\documentclass[prb,reprint,superscriptaddress,showkeys,twocolumn,amsmath,amssymb]{revtex4-2}
\usepackage{graphicx}
\usepackage[colorlinks=true,allcolors=blue]{hyperref}
\usepackage{braket}

\begin{document}

\title{\large{Discrete control of capacitance in quantum circuits}}
\author{R.-P. Riwar}
\affiliation{Peter Gr\"unberg Institute, Theoretical Nanoelectronics, Forschungszentrum J\"ulich, D-52425 J\"ulich, Germany}

\begin{abstract}

Precise in-situ control of system parameters is indispensable for all quantum hardware applications. The capacitance in a circuit, however, is usually a simple consequence of electrostatics, and thus quite literally cast in stone. We here propose a way to control the charging energy of a given island by exploiting recently predicted Chern insulator physics in common Cooper-pair transistors, where the capacitance switches between discrete values given by the Chern number. We identify conditions for which the discrete control benefits from exponentially reduced noise sensitivity to implement protected tunable qubits.

\end{abstract}

\maketitle


\textit{Introduction} -- It is one of the main cruxes of quantum hardware that in-situ tuning of system parameters invariably comes at the cost of exposing the system to noise from the corresponding control knobs, such as, for instance, in flux-tunable Josephson junctions~\cite{Koch_1983,Wellstood_1987,Vion_2002,Barends_2013,Yan_2016,Quintana_2017,Valery_2022}. The appeal of topological phase transitions in quantum systems~\cite{Sarma:2015aa,Bansil:2016aa} is therefore obvious: to store the quantum information in a protected subspace that is insensitive to small continuous variations. This idea comes with its own subset of challenges. Qubits comprised of solid state materials with topological properties in their band structure~\cite{Hatsugai:1993aa,Kitaev2001,Kane2005TopologicalOrder,Bernevig:2006aa,Fu:2007aa,Murakami:2007aa,Fu:2008aa,Prodan:2010aa,Hasan:2010aa,Qi:2011aa,Wan:2011aa,Burkov:2011aa,Meng:2012aa,Alicea:2012aa,Hosur:2013aa,Yang:2014aa,Bednik:2015aa,Lv:2015aa,Chiu:2016aa,Sato:2017aa,Bernevig:2018aa,Armitage:2018aa,Lutchyn:2018aa,Peralta-Gavensky:2019aa,Sakurai:2020aa,rui2020higherorder} are to this day challenging to implement and to tune.
Recently, an alternative idea gained traction: to explore topological properties not in the bandstructure of materials, but in the emergent transport degrees of freedom in circuits, where they seem to be much more abundant~\cite{Leone2008,Leone_2013,Yokoyama2015TopolABSmultitJJ,Riwar2016,Strambini_2016,Eriksson2017,Meyer:2017aa,Deb:2018aa,Riwar_2019b,Repin:2019aa,Repin_2020,Fatemi_2020,Peyruchat_2020,Klees:2020aa,Klees_2021,Weisbrich_2021,Weisbrich_2021_monopoles,Chirolli_2021,Herrig_2022,Melo2022,Javed_2023}. For this type of topological effect, however, concrete pathways towards a protected qubit have not yet been explored much.

\begin{figure}[h!]
    \centering
    \includegraphics[width=0.9\columnwidth]{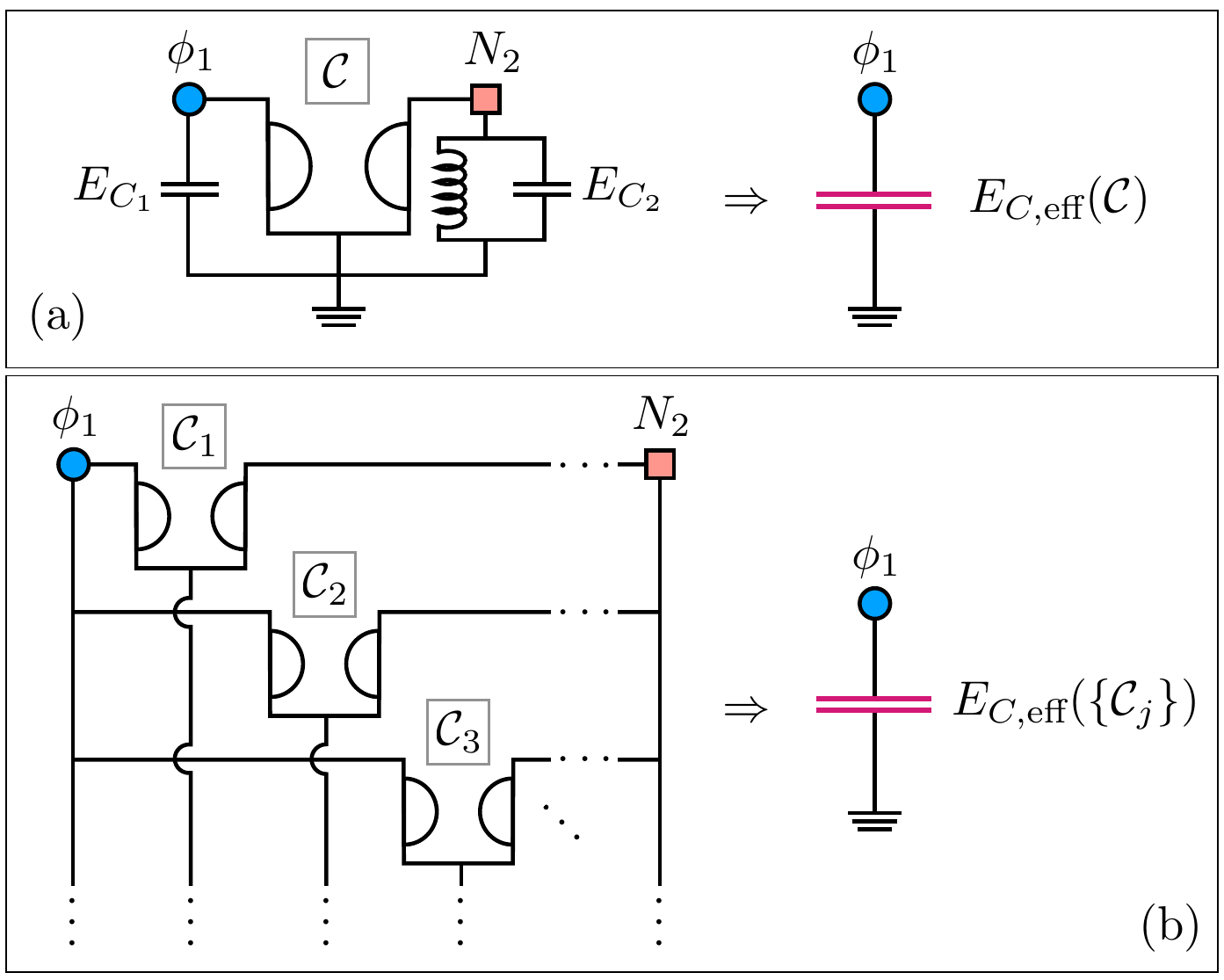}
    \caption{The principle of a topologically tunable capacitance. The Chern physics of regular transistors, here represented by Tellegen's symbol as a type of nonreciprocal element, implement a charging energy tunable by means of the Chern number $\mathcal{C}$ (a). By shunting several nonreciprocal elements in parallel (b), tuning between an exponentially increasing number of possible charging energy values can be achieved. }
    \label{fig:digital_capacitance}
\end{figure}

Instead of directly building a qubit, we here introduce the concept of utilizing topological transitions in a conventional circuit element to implement a protected control of a system parameter. Surprisingly, the parameter in question turns out to be the capacitance of a given charge island (see Fig.~\ref{fig:digital_capacitance}), which was so far not expected to be tunable. To accomplish this feat we exploit the recently predicted Chern physics of conventional Cooper-pair transistors~\cite{Herrig_2022}. Flexible tuning of the capacitance might have a variety of applications, for instance to probe recently proposed quasiperiodic effects in charge space~\cite{herrig2022quasiperiodic} which have the potential to emulate a transport versions of twistronics~\cite{GeimGrigorieva2913,CaoJarillo,GonzalezCirac2019,FuPixley2020,SalamonRakshit2020,ChouPixley2020,MaoSenthil2021,MengZhang2021,LeePixley2022}. We here explicitly explore the necessary conditions to harness the topological protection to build a topologically tunable transmon. 

For a regular, linear capacitor, the charging energy is simply given by the energy stored in the electric field built up when separating charges and is inversely proportional to the capacitance. The last years and decades have however seen remarkable progress in coming up with more general notions of capacitances. Ferroelectric materials have been intensively studied~\cite{Landauer_1976,Catalan_2015,Ng_2017,Hoffmann_2018,Lukyanchuk_2019,Hoffmann_2020} and shown to give rise to a highly nonlinear charge-voltage relationship. Such materials are for instance of interest to create a lever-arm effect in field-effect transistors, by implementing partially negative capacitances~\cite{Salahuddin_2008}. The quantum capacitance~\cite{Luryi_1988} in certain lower dimensional systems has also been shown to be negative~\cite{Wang_2013,Choi_2016}. In superconducting circuits, quantum phase slip junctions~\cite{Giordano1988,Bezryadin2000,Lau_2001,Buchler2004,Mooij_2005,Mooij_2006,Arutyunov2008,Astafiev_2012,Ulrich_2016,deGraaf_2018,Li_2019,Shaikhaidarov_2022} can under certain assumptions~\cite{koliofoti2022} be regarded as nonlinear capacitors. Time varying fluxes may give rise to effectively negative or time-dependent junction capacitances~\cite{Riwar_2022}. Very recently, the notion of nonlinear capacitance has been strongly generalized to include quasiperiodicity in charge space~\cite{herrig2022quasiperiodic}, for which the here proposed effect could be pivotal, as it would allow to tune the quasiperiodicity parameter. 

We examine two important sources of perturbations. On the one hand, the transistors may give rise to a remaining leakage Josephson effect. We expect however that the detrimental effect of fluctuations of the critical current can significantly mitigated when using gate-controlled junctions~\cite{Larsen_2015,deLange_2015,Kjaergaard_2017,Casparis_2018,Lee_2019,Graziano_2022}, which can in principle be pinched off to a regime, where the Josephson effect \textit{and} its fluctuations are exponentially suppressed. On the other hand, the capacitance control comes with increased offset charge noise sensitivity due to the extra coupled island. This effect should however be less relevant when used for a transmon~\cite{Koch_2007}. The extra added offset charge sensitivity is nonetheless of fundamental interest, as it illustrates the importance of compact variables, an issue which has recently seen a revival within the community~\cite{Likharev_1985,Loss_1991,Koch_2009,Mizel_2020,Thanh_2020,Murani_2020,Hakonen_2021,Murani_2021,Roman2021,Devoret_2021,Kaur_2021,Kenawy_2022,Masuki_2022,kuzmin2023observation,kashuba2023counting}.

\begin{figure}
    \centering
    \includegraphics[width=\columnwidth]{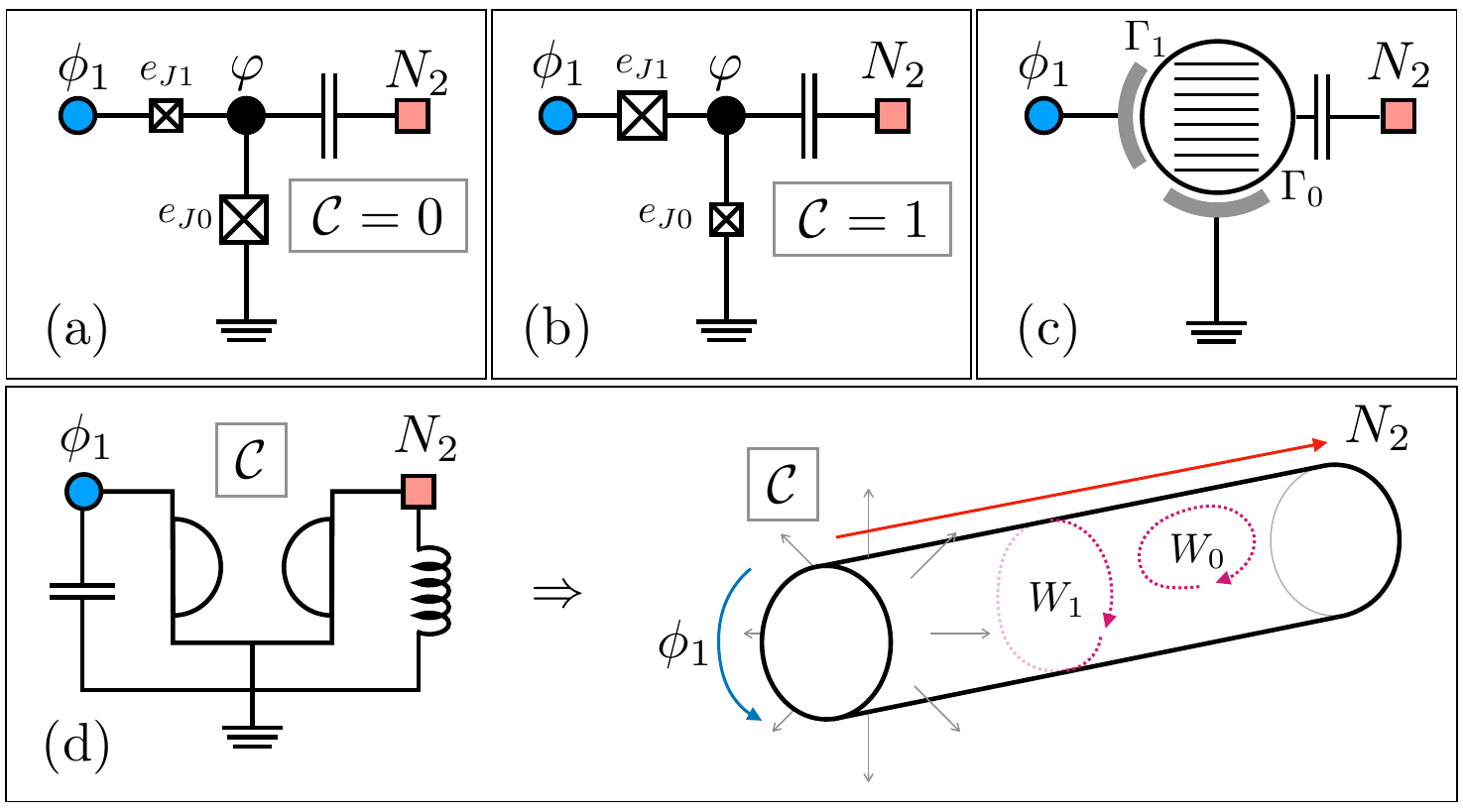}
    \caption{Possible physical realizations of the transistor, either by means of a central charge island (a,b) or a quantum dot with discrete levels (c). The value of the Chern number $\mathcal{C}$ depends on which of the two junctions is dominant (either described by the Josephson energies $e_{J0}$ and $e_{J1}$ in a and b, or by the Cooper pair tunneling rates $\Gamma_0$ and $\Gamma_1$ in c). The integrated circuit realizes the equivalent of a particle moving on a cylinder, where the Chern number is the equivalent of the magnetic flux (d). On a cylinder there are two distinct types of gauge invariant loops ($W_0$ and $W_1$) -- an important detail for the correct description of offset charge noise. }
    \label{fig:chern_cylinder}
\end{figure}

\textit{Transistors and Chern numbers} -- The here proposed effect relies on the recently studied~\cite{Herrig_2022} topological properties of regular Cooper-pair transistors. While Ref.~\cite{Herrig_2022} considered a specific example of a transistor with a central superconducting charge island coupled by means of regular SIS Josephson junctions (see Fig.~\ref{fig:chern_cylinder}a,b), we insist that the effect is highly generic, and emerges for a wide class of transistor realizations. In Fig.~\ref{fig:chern_cylinder}c we show for concreteness one possible alternative, where the transistor is made out of a quantum dot. The two example models shown in Fig.~\ref{fig:chern_cylinder} can be regarded as the same device in complementary regimes: in the first, many-body interactions on the central transistor island are taken into account, but the level spacing due to orbital effects is neglected. In the second, interactions are absent (assuming that electrons spend too little time inside the quantum dot to interact), but the small dot size results in a level spacing. Both models, including their ground state properties, are discussed in more detail in the supplementary material.

For the central principle to work, the following two ingredients are required. i) A transistor with three contacts: one to ground, and two contacts that we refer to as node $1$ and node $2$. Ground and node $1$ are electrically coupled, and exchange Cooper pairs across the transistor. The transistor gate is denoted as node $2$, and is only capacitively coupled. The two only relevant degrees of freedom outside the transistor are therefore the charge $N_2$ on the transistor gate, as well as the superconducting phase $\phi_1$ on node $1$. The physics of the transistor are therefore characterized by the Hamiltonian
\begin{equation}
    H_\text{T}(\phi_1,N_2)=\sum_m \epsilon_m\left\vert m(\phi_1,N_2\right\rangle \left\langle m(\phi_1,N_2)\right\vert\ ,
\end{equation}
ii) The transistor needs to have a tunable asymmetry for the electric coupling to ground and node $1$, and a gapped spectrum with a nondegenerate ground state. Thanks to the gap, the physics of the transistor at low energies can be captured entirely by the ground state ($m=0$) energy $\epsilon=\epsilon_0(\phi_1,N_2)$ and the corresponding Berry curvature $\mathcal{B}=-2\text{Im}\left[\langle\partial_{\phi_1}0\vert\partial_{N_2}0\rangle\right]$.

As shown in Ref.~\cite{Herrig_2022}, 
the Chern number $\mathcal{C}=\iint_{\text{BZ}}d\phi_1dN_2 \mathcal{B}/2\pi$ depends on the transistor asymmetry. It assumes the values $0$ or $1$, depending on whether Cooper-pair tunneling is more likely to occur to ground (for the explicit examples in Fig.~\ref{fig:chern_cylinder}, $e_{J0}>e_{J1}$, respectively $\Gamma_0>\Gamma_1$) or to node $1$ ($e_{J0}<e_{J1}$, respectively $\Gamma_0<\Gamma_1$). The Brillouin zone ($\text{BZ}$) goes from $0$ to $2\pi$ for $\phi_1$ and from $0$ to $1$ for $N_2$~\footnote{This is true if $N_2$ is rescaled such that a change of $N_2$ by $\pm 1$ corresponds to an induced offset charge of an individual Cooper pair on the transistor.}. The Chern number can be directly measured by time-dependently driving both $\phi_1$ and $N_2$, yielding a quantized dc current response into contact $\phi_1$, of the form $I_1=2e\mathcal{C}\dot{N}_2$. In short, the dc part of the displacement current induced by $\dot{N}_2$ flows exclusively through the electric contact with stronger tunnel coupling -- irrespective of how much stronger it is. 

\textit{Mechanism for capacitance control} -- Let us now integrate the transistor into a larger circuit, and predict the emergent behaviour. 
Both nodes shall be connected to capacitors (described by the charging energies $E_{C_1}$ and $E_{C_2}$) and node $2$ in addition to an inductor ($E_{L_2}$), see Fig.~\ref{fig:digital_capacitance}(a). The resulting Hamiltonian reads
\begin{equation}
    H=H_\text{T}(\widehat{\phi}_1,\widehat{N}_2)+E_{C_1}(\widehat{N}_1+N_{g1})^2+E_{C_2}\widehat{N}_2^2+E_{L_2}\widehat{\phi}_2^2\ ,
\end{equation}
with $[\widehat{\phi}_j,\widehat{N}_{j^\prime}]=i\delta_{jj^\prime}$. We include an additional gate applied to node $1$, giving rise to the induced offset charge $N_{g1}$ (not explicitly shown in the figures).
Taking into account the aforementioned energy gap in the transistor, the problem can be simplified in two steps. In a first step, we transform away the $(\phi_1,N_2)$-dependence in the basis of $H_T$, by the unitary $U=\sum_m\vert \widetilde{m}\rangle\langle m(\phi_1,N_2)\vert$, where the new basis $\vert \widetilde{m}\rangle$ is explicitly constant in both $\phi_1$ and $N_2$. The node operators thus transform as $N_1\rightarrow N_1-iU\partial_{\phi_1}U^\dagger$ and $\phi_2\rightarrow \phi_2+iU\partial_{N_2}U^\dagger$. In a second step, we project away all but the ground state $m=0$.
The Hamiltonian then reduces to
\begin{equation}\label{eq:H_cylinder}
\begin{split}
    H=E_{C_1}\left[\widehat{N}_1+N_{g1}+\mathcal{A}_1\left(\widehat{\phi}_1,\widehat{N}_2\right)\right]^2+E_{C_2}\widehat{N}_2^2 \\+ E_{L_2}\left[\widehat{\phi}_2+\mathcal{A}_2\left(\widehat{\phi}_1,\widehat{N}_2\right)\right]^2 +\epsilon\left(\widehat{\phi}_1,\widehat{N}_2\right)\ ,
\end{split}
\end{equation}
where $\mathcal{A}_{1}=-i\langle 0\vert\partial_{\phi_1}\vert 0 \rangle$ and $\mathcal{A}_{2}=i\langle 0\vert\partial_{N_2}\vert 0 \rangle$ are the Berry connections satisfying
\begin{equation}\label{eq:berry_connection}
    \mathcal{B}=\partial_{N_2}\mathcal{A}_1+\partial_{\phi_1}\mathcal{A}_2\ .
\end{equation}
Importantly, note that the curl in Eq.~\eqref{eq:berry_connection} is defined with a $+$ sign instead of the usual $-$ sign, due to the conjugate nature of the charge ($N_2$) and phase ($\phi_1$) variable.
The quantities $\epsilon$, $\mathcal{A}_1$, and $\mathcal{A}_2$ therefore play a role equivalent to the electromagnetic scalar and vector potentials.
Consequently, the Berry curvature is the equivalent of a magnetic field acting on a particle moving in $(\phi_1,N_2)$-space, whose total flux per Brillouin zone is $\mathcal{C}$. We notice that the functioning of the transistor is akin to a gyrator~\cite{Viola_2014,Rymarz2021}, except that the latter gives rise to a magnetic field in $(\phi_1,\phi_2)$-space instead of $(\phi_1,N_2)$-space. This prompts us to refer to the transistor as a half-dual nonreciprocal circuit element. This ``half-duality'' is captured in Figs.~\ref{fig:digital_capacitance} and~\ref{fig:chern_cylinder} by denoting the phase node with a circle, and the charge node with a square.

There is one further noteworthy specialty. Since $\phi_1$ space is compact (i.e., the charge on node $1$ can only change by integer Cooper-pairs), our system realizes the equivalent of particle moving on a cylinder, see Fig.~\ref{fig:chern_cylinder}(d). The topology of the base manifold has an influence on the gauge degrees of freedom for the Berry connections: on a cylinder, there are two types of topologically distinct loops along which gauge invariant phases can be picked up. $W_0$ is the regular type of loop which exists also on a flat extended space, and does not probe the cylinder topology. The requirement that the phase picked up on such loops be gauge-invariant is captured by Eq.~\eqref{eq:berry_connection}. The second type of loop, $W_1$, owes its existence to the genus $1$ of the cylinder manifold, and provides an additional constraint on the vector potentials. Namely, the phase picked up along such loops is likewise gauge-invariant. Due to Eq.~\eqref{eq:berry_connection}, different $W_1$ loops are not independent. Therefore, we can choose one simple loop along $\phi_1$, say, by fixing $N_2=0$. The phase $e^{i\int_0^2\pi d\phi_1 \mathcal{A}_1}$ for $N_2=0$ then needs to be fixed by an extra condition. To determine it, one has to compute the corresponding Berry phase picked up, when parallel-transporting the ground state of the transistor Hamiltonian $H_\text{T}$ along this path. For the main results of this work, this particular detail is irrelevant. But it does matter to correctly estimate the influence of offset charge noise, as we point out at the end of this work.

For simplicity, let us first discuss the case when $\epsilon$ and $\mathcal{B}$ are independent of $\phi_1$ and $N_2$ (flat energy bands with flat curvature). The scalar potential $\epsilon$ is then an irrelevant constant, and the equivalent magnetic field is homogeneous on the entire cylinder, given by $\mathcal{B}=\mathcal{C}$. In order to satisfy Eq.~\eqref{eq:berry_connection}, we choose the equivalent of the Landau gauge $\mathcal{A}_1=\mathcal{C}N_2$ and $\mathcal{A}_2=0$. For $E_{C_2}=0$, the Hamiltonian $H$ in Eq.~\eqref{eq:H_cylinder} would now provide regular Landau levels. It is however not physical that the charge on node $2$ can be accumulated indefinitely, hence the presence of a finite $E_{C_2}$. Including a nonzero $E_{C_2}$, the Hamiltonian $H$ is diagonalized as
\begin{equation}\label{eq:H_chern}
    H=\underbrace{\frac{1}{\frac{1}{E_{C_1}}+\frac{1}{E_{C_2}}\mathcal{C}^2 }}_{\equiv E_{C,\text{eff}}(\mathcal{C})}(\widehat{N}_1+N_{g1})^2+\omega_{\text{LC}}\widehat{b}^\dagger\widehat{b}\ ,
\end{equation}
where $\omega_\text{LC}=2\sqrt{E_{L}(\mathcal{C}^2E_{C_1}+E_{C_2})}$ is the eigenfrequency of the LC resonator coupled to node 2, whose eigenmodes are annihilated with the ladder operator $\widehat{b}=\gamma\widehat{N}_1/2-i\widehat{\phi}_2/\gamma$ and $\gamma=\sqrt{\omega_\text{LC}/E_{L}}$. 

The above constitutes our central result: the Chern physics of a generic transistor, if coupled to an additional charge island and an LC resonator, reduce to an effective charge island with a charging energy that is discretely tunable by the Chern number. 
To increase the tuning capability, we further note that nothing stops us from coupling the two nodes with several transistors, see Fig.~\ref{fig:digital_capacitance}(b), each with their own Chern number $\mathcal{C}_j$. Given the above framework, it is now a straightforward exercise to show that the resulting charging energy becomes
\begin{equation}
    E_{C,\text{eff}}(\{\mathcal{C}_j\})=\frac{1}{\frac{1}{E_{C_1}}+\frac{\left(\sum_j\mathcal{C}_j\right)^2}{E_{C_2}}}\ .
\end{equation}
Importantly, the number of values the charging energy can take grows exponentially with the total number $J$ of transistors, as $\sim 2^J$. 
Instead of connecting all transistors to the same node $2$ with just one LC resonator, one can potentially increase the flexibility and range of the topological tuning, by increasing the number of nodes on the right hand side of Fig.~\ref{fig:digital_capacitance}(b), that is, to connect each transistor node $j$ to a separate LC resonator, $N_2\rightarrow N_{2,j}$, each with their own charging energy $E_{C_2,j}$. Of course, while this increases the tuning range, it adds extra hardware overhead through the additional LC resonators. The pros and cons would thus have to be weighed against one another for a concrete experimental realization.

\textit{Applications and perturbations} -- To explore the applicability of the here proposed topological capacitive control we have to identify the leading perturbations when going beyond the above idealistic assumptions. For simplicity, we return to the single transistor device as in Fig.~\ref{fig:digital_capacitance}a. The main simplifying assumption we made above is to take both $\epsilon$ and $\mathcal{B}$ to be flat. We therefore update our calculation by taking into account deviation from flat bands perturbatively. We do so by assuming that $E_{L_2}$ is large, such that we may integrate out the LC resonator degree of freedom. In this regime, it turns out that only a finite $\phi_1$-dependence is potentially harmful. To appreciate this fact, consider for instance the impact of a finite $N_2$-dependence on $\epsilon$: integrating out the LC resonator, the function $\epsilon(N_2)$ will be averaged over $N_2$ with the weight given by the LC ground state wave function, and thus contribute only a constant offset -- which is obviously inconsequential. We therefore consider $\epsilon=\epsilon_0+\delta\epsilon(\phi_1)$ (where $\epsilon_0$ is a discardable constant) and $\mathcal{B}=\mathcal{C}+\delta\mathcal{B}(\phi_1)$, with $\delta\epsilon\ll\omega_\text{LC}$ and $\delta\mathcal{B}\ll 1$. We now find  $\mathcal{A}_1=(\mathcal{C}+\delta\mathcal{B})N_2$ whereas $\mathcal{A}_2$ remains 0. By means of a Schrieffer-Wolff transformation, mapping onto the ground state of the LC resonator, we get in leading order (details shown in the supplementary material)
\begin{equation}\label{eq:leakage}
    H\approx E_{C,\text{eff}}\left(\mathcal{C}\right)\left(\widehat{N}_{1}+N_{g1}\right)^{2}+\underbrace{\delta\epsilon+\frac{\mathcal{C}}{2}E_{C,\text{eff}}\left(\mathcal{C}\right)\frac{\omega_{\text{LC}}}{E_{C_{2}}}\delta\mathcal{B}}_{\equiv E_{J,\text{leak}}(\phi_1)}\ ,
\end{equation}
where the correction terms due to deviations from the flat band approximation represent a leakage Josephson effect (due to their $\phi_1$-dependence). Since $\omega_\text{LC}$ is assumed to be large, it has to be evaluated for the specific model (and parameter regime) at hand, whether the first term ($\sim\delta\epsilon$) or the second term ($\sim\delta\mathcal{B}$) dominates.

In order to mitigate potential ramifications from this perturbation, it is important to identify the main physical cause of a finite $\phi_1$-dependence of $\epsilon$ and $\mathcal{B}$. For the two example realizations of the transistor in Fig.~\ref{fig:chern_cylinder}, it can be shown (supplementary) that both $\delta\epsilon$ and $\delta\mathcal{B}$ scale with the weaker of the two junction energies, that is $\delta\epsilon,\delta\mathcal{B}\sim e_{J0},\Gamma_0$ ($\sim e_{J1},\Gamma_1$) for $\mathcal{C}=1$ ($\mathcal{C}=0$). Since these junctions are tunable, they will be susceptible to external noise on the corresponding control knobs, such that the leakage Josephson effect is at risk of being noisy, and may thus thwart our idea of exploiting the topologically protected tuning of the charging energy. For instance, suppose we want to exploit the discrete capacitive control for the implementation of a tunable transmon. We would then couple node $1$ with a nontunable Josephson junction with energy $E_J\gg E_{C,\text{eff}}$, such that the qubit frequency is approximated as $\sim\sqrt{E_{C,\text{eff}}E_J}$. But crucially, the noisy leakage Josephson effect from the transistor would have to be added to $E_J$, such that our more complicated device would not benefit from any advantage over a simpler transmon where the Josephson junction $E_J$ itself is tunable (and thus noisy). But there is a very promising way out, depending on how the control of the junction energy is realized.

The most common control of the Josephson junction energy is via a dc-SQUID~\cite{Koch_1983}, which is however known to only have a sweet spot of quadratically suppressed flux-noise sensitivity when the Josephson energy is maximal. For our purposes however, we require a sweet spot in the regime where the junction is \textit{weak}. As foreshadowed already in the introduction, controlling the transparency of a junction by means of electrical gate modulations along similar lines as in Refs.~\cite{Larsen_2015,deLange_2015,Kjaergaard_2017,Casparis_2018,Lee_2019,Graziano_2022} might hold enormous potential. Namely, the junction can in principle be pinched off and tuned to zero coupling -- up to exponentially suppressed tunneling processes. While this is not commonly where these junctions are operated, for our purposes, the likewise exponentially suppressed noise sensitivity on the corresponding gate is exactly what we are are after. Namely, when tuning from $\mathcal{C}=0$ to $\mathcal{C}=1$ in the transistor, the weaker junction can be put to the sweet spot of almost zero tunneling, while the influence of the stronger junction (which is generally in a noisy regime) is banned to high energies, where it does not interfere with the low-energy physics.

After these applied considerations, let us finish on a more fundamental note, with a word of caution regarding offset charge fluctuations. The above Eqs.~\eqref{eq:H_chern} and~\eqref{eq:leakage} might invite the wrong impression that the amplitude of the offset charge noise remains the same independent of the value of $\mathcal{C}$, i.e., it is simply given by the amplitude of fluctuations of $N_{g1}$. This is however not true, as can be easily seen when adding offset charges to other parts of the device. In particular, for $\mathcal{C}=1$, the islands of node $1$ and $2$ are effectively joined to a single island, such that offset charge noise of the combined island is given as the sum of fluctuations coupling to node $1$ and $2$. We can understand this effect in two complementary ways. Including an additional offset charge to the charging energy term $\sim E_{C_2}$ in node $2$ in Eq.~\eqref{eq:H_cylinder}, $\widehat{N}_2\rightarrow \widehat{N}_2+N_{g2}$, we see that the resulting offset charge in Eqs.~\eqref{eq:H_chern} and~\eqref{eq:leakage} must be shifted by $N_{g1}\rightarrow N_{g1}+\mathcal{C}N_{g2}$. Alternatively, by means of a unitary transformation, we may eliminate $N_{g2}$ inside the charging energy term $\sim E_{C_2}$, and instead have it appear inside $H_\text{T}$. Now, it would seem that $N_\text{g2}$ has disappeared from the problem -- giving rise to a serious inconsistency. Such a conclusion would however be due to a naive interpretation of gauge invariance, in particular neglecting the need to fix the path integral of the vector potential along paths of type $W_1$ (Fig.~\ref{fig:chern_cylinder}d). The additional contribution $+\mathcal{C}N_{g2}$ to the offset charge is hidden in exactly this condition. Correctly fixing the phase along $W_1$ leads to a shift in the vector potential, from $\mathcal{A}_1=\mathcal{C}N_2$ to $\mathcal{A}_1=\mathcal{C}(N_2+N_{g2})$. While this leaves the Berry curvature invariant, as per Eq.~\eqref{eq:berry_connection}, it nonetheless changes the resulting energy spectrum in Eqs.~\eqref{eq:H_chern} and~\eqref{eq:leakage}.

We summarize that it is of importance to correctly apply gauge transformations on a compact manifold (here, the cylinder), to avoid inconsistent predictions, or missing important contributions to noise sources which affect, e.g., the system's relaxation rates. This caveat serves as a comment on a currently ongoing debate within the quantum circuit community about the relevance of compactness of circuit degrees of freedom~\cite{Likharev_1985,Loss_1991,Koch_2009,Mizel_2020,Thanh_2020,Murani_2020,Hakonen_2021,Murani_2021,Roman2021,Devoret_2021,Kaur_2021,Kenawy_2022,Masuki_2022,kuzmin2023observation,kashuba2023counting}. The debate can be formulated in the framework of gauge degrees of freedom of quantum systems on compact manifolds as follows. The above extra condition on the vector potential due to loops of $W_1$ is equivalent to demanding that the wave function on the cylinder be single-valued (equivalent to compact $\phi_1$). Disregarding it is equivalent to demanding that only the modulus square of the wave function (the probability density of the quantum system) be single-valued (noncompact $\phi_1$). Proponents of the latter point of view would likely argue that one could just redefine $N_{g1}+\mathcal{C}N_{g2}$ as a new $N_{g1}^\prime$, and still get the formally correct Hamiltonian. However, as the above argument illustrates for our concrete model, this approach harbors the danger of unnecessarily blurring the microscopic origin of charge fluctuations, and may fail to provide the correct quantitative predictions for qubit relaxation (and potentially other) processes.

\textit{Conclusions and outlook} -- By exploiting recently predicted Chern physics of conventional Cooper-pair transistors, we introduce the concept of a topologically tunable capacitance for superconducting circuits. We identify optimal parameter regimes for the effect to be observed and to be used for tunable qubits with reduced sensitivity from external noise. We expect that the here proposed principle may provide an interesting alternative towards protected quantum hardware, where the considered topological effect is not directly used to build a qubit with, e.g., a protected ground state degeneracy, but instead to have a precise, discrete control of a given tuning parameter.

\textit{Acknowledgements} -- 	We warmly thank O. Kashuba and G. Catelani for fruitful discussions. This work has been funded by the German Federal Ministry of Education and Research within the funding program Photonic Research Germany under the contract number 13N14891.
\bibliographystyle{apsrev4-2}
\bibliography{paper1,paper2,references,biblio}

\begin{thebibliography}{115}%
\makeatletter
\providecommand \@ifxundefined [1]{%
 \@ifx{#1\undefined}
}%
\providecommand \@ifnum [1]{%
 \ifnum #1\expandafter \@firstoftwo
 \else \expandafter \@secondoftwo
 \fi
}%
\providecommand \@ifx [1]{%
 \ifx #1\expandafter \@firstoftwo
 \else \expandafter \@secondoftwo
 \fi
}%
\providecommand \natexlab [1]{#1}%
\providecommand \enquote  [1]{``#1''}%
\providecommand \bibnamefont  [1]{#1}%
\providecommand \bibfnamefont [1]{#1}%
\providecommand \citenamefont [1]{#1}%
\providecommand \href@noop [0]{\@secondoftwo}%
\providecommand \href [0]{\begingroup \@sanitize@url \@href}%
\providecommand \@href[1]{\@@startlink{#1}\@@href}%
\providecommand \@@href[1]{\endgroup#1\@@endlink}%
\providecommand \@sanitize@url [0]{\catcode `\\12\catcode `\$12\catcode
  `\&12\catcode `\#12\catcode `\^12\catcode `\_12\catcode `\%12\relax}%
\providecommand \@@startlink[1]{}%
\providecommand \@@endlink[0]{}%
\providecommand \url  [0]{\begingroup\@sanitize@url \@url }%
\providecommand \@url [1]{\endgroup\@href {#1}{\urlprefix }}%
\providecommand \urlprefix  [0]{URL }%
\providecommand \Eprint [0]{\href }%
\providecommand \doibase [0]{https://doi.org/}%
\providecommand \selectlanguage [0]{\@gobble}%
\providecommand \bibinfo  [0]{\@secondoftwo}%
\providecommand \bibfield  [0]{\@secondoftwo}%
\providecommand \translation [1]{[#1]}%
\providecommand \BibitemOpen [0]{}%
\providecommand \bibitemStop [0]{}%
\providecommand \bibitemNoStop [0]{.\EOS\space}%
\providecommand \EOS [0]{\spacefactor3000\relax}%
\providecommand \BibitemShut  [1]{\csname bibitem#1\endcsname}%
\let\auto@bib@innerbib\@empty
\bibitem [{\citenamefont {Koch}\ \emph {et~al.}(1983)\citenamefont {Koch},
  \citenamefont {Clarke}, \citenamefont {Goubau}, \citenamefont {Martinis},
  \citenamefont {Pegrum},\ and\ \citenamefont {van Harlingen}}]{Koch_1983}%
  \BibitemOpen
  \bibfield  {author} {\bibinfo {author} {\bibfnamefont {R.~H.}\ \bibnamefont
  {Koch}}, \bibinfo {author} {\bibfnamefont {J.}~\bibnamefont {Clarke}},
  \bibinfo {author} {\bibfnamefont {W.~M.}\ \bibnamefont {Goubau}}, \bibinfo
  {author} {\bibfnamefont {J.~M.}\ \bibnamefont {Martinis}}, \bibinfo {author}
  {\bibfnamefont {C.~M.}\ \bibnamefont {Pegrum}},\ and\ \bibinfo {author}
  {\bibfnamefont {D.~J.}\ \bibnamefont {van Harlingen}},\ }\href
  {https://doi.org/10.1007/BF00683423} {\bibfield  {journal} {\bibinfo
  {journal} {Journal of Low Temperature Physics}\ }\textbf {\bibinfo {volume}
  {51}},\ \bibinfo {pages} {207} (\bibinfo {year} {1983})}\BibitemShut
  {NoStop}%
\bibitem [{\citenamefont {Wellstood}\ \emph {et~al.}(1987)\citenamefont
  {Wellstood}, \citenamefont {Urbina},\ and\ \citenamefont
  {Clarke}}]{Wellstood_1987}%
  \BibitemOpen
  \bibfield  {author} {\bibinfo {author} {\bibfnamefont {F.~C.}\ \bibnamefont
  {Wellstood}}, \bibinfo {author} {\bibfnamefont {C.}~\bibnamefont {Urbina}},\
  and\ \bibinfo {author} {\bibfnamefont {J.}~\bibnamefont {Clarke}},\ }\href
  {https://doi.org/10.1063/1.98041} {\bibfield  {journal} {\bibinfo  {journal}
  {Applied Physics Letters}\ }\textbf {\bibinfo {volume} {50}},\ \bibinfo
  {pages} {772} (\bibinfo {year} {1987})}\BibitemShut {NoStop}%
\bibitem [{\citenamefont {Vion}\ \emph {et~al.}(2002)\citenamefont {Vion},
  \citenamefont {Aassime}, \citenamefont {Cottet}, \citenamefont {Joyez},
  \citenamefont {Pothier}, \citenamefont {Urbina}, \citenamefont {Esteve},\
  and\ \citenamefont {Devoret}}]{Vion_2002}%
  \BibitemOpen
  \bibfield  {author} {\bibinfo {author} {\bibfnamefont {D.}~\bibnamefont
  {Vion}}, \bibinfo {author} {\bibfnamefont {A.}~\bibnamefont {Aassime}},
  \bibinfo {author} {\bibfnamefont {A.}~\bibnamefont {Cottet}}, \bibinfo
  {author} {\bibfnamefont {P.}~\bibnamefont {Joyez}}, \bibinfo {author}
  {\bibfnamefont {H.}~\bibnamefont {Pothier}}, \bibinfo {author} {\bibfnamefont
  {C.}~\bibnamefont {Urbina}}, \bibinfo {author} {\bibfnamefont
  {D.}~\bibnamefont {Esteve}},\ and\ \bibinfo {author} {\bibfnamefont {M.~H.}\
  \bibnamefont {Devoret}},\ }\href {https://doi.org/10.1126/science.1069372}
  {\bibfield  {journal} {\bibinfo  {journal} {Science}\ }\textbf {\bibinfo
  {volume} {296}},\ \bibinfo {pages} {886} (\bibinfo {year}
  {2002})}\BibitemShut {NoStop}%
\bibitem [{\citenamefont {Barends}\ \emph {et~al.}(2013)\citenamefont
  {Barends}, \citenamefont {Kelly}, \citenamefont {Megrant}, \citenamefont
  {Sank}, \citenamefont {Jeffrey}, \citenamefont {Chen}, \citenamefont {Yin},
  \citenamefont {Chiaro}, \citenamefont {Mutus}, \citenamefont {Neill},
  \citenamefont {O'Malley}, \citenamefont {Roushan}, \citenamefont {Wenner},
  \citenamefont {White}, \citenamefont {Cleland},\ and\ \citenamefont
  {Martinis}}]{Barends_2013}%
  \BibitemOpen
  \bibfield  {author} {\bibinfo {author} {\bibfnamefont {R.}~\bibnamefont
  {Barends}}, \bibinfo {author} {\bibfnamefont {J.}~\bibnamefont {Kelly}},
  \bibinfo {author} {\bibfnamefont {A.}~\bibnamefont {Megrant}}, \bibinfo
  {author} {\bibfnamefont {D.}~\bibnamefont {Sank}}, \bibinfo {author}
  {\bibfnamefont {E.}~\bibnamefont {Jeffrey}}, \bibinfo {author} {\bibfnamefont
  {Y.}~\bibnamefont {Chen}}, \bibinfo {author} {\bibfnamefont {Y.}~\bibnamefont
  {Yin}}, \bibinfo {author} {\bibfnamefont {B.}~\bibnamefont {Chiaro}},
  \bibinfo {author} {\bibfnamefont {J.}~\bibnamefont {Mutus}}, \bibinfo
  {author} {\bibfnamefont {C.}~\bibnamefont {Neill}}, \bibinfo {author}
  {\bibfnamefont {P.}~\bibnamefont {O'Malley}}, \bibinfo {author}
  {\bibfnamefont {P.}~\bibnamefont {Roushan}}, \bibinfo {author} {\bibfnamefont
  {J.}~\bibnamefont {Wenner}}, \bibinfo {author} {\bibfnamefont {T.~C.}\
  \bibnamefont {White}}, \bibinfo {author} {\bibfnamefont {A.~N.}\ \bibnamefont
  {Cleland}},\ and\ \bibinfo {author} {\bibfnamefont {J.~M.}\ \bibnamefont
  {Martinis}},\ }\href {https://doi.org/10.1103/PhysRevLett.111.080502}
  {\bibfield  {journal} {\bibinfo  {journal} {Phys. Rev. Lett.}\ }\textbf
  {\bibinfo {volume} {111}},\ \bibinfo {pages} {080502} (\bibinfo {year}
  {2013})}\BibitemShut {NoStop}%
\bibitem [{\citenamefont {Yan}\ \emph {et~al.}(2016)\citenamefont {Yan},
  \citenamefont {Gustavsson}, \citenamefont {Kamal}, \citenamefont {Birenbaum},
  \citenamefont {Sears}, \citenamefont {Hover}, \citenamefont {Gudmundsen},
  \citenamefont {Rosenberg}, \citenamefont {Samach}, \citenamefont {Weber},
  \citenamefont {Yoder}, \citenamefont {Orlando}, \citenamefont {Clarke},
  \citenamefont {Kerman},\ and\ \citenamefont {Oliver}}]{Yan_2016}%
  \BibitemOpen
  \bibfield  {author} {\bibinfo {author} {\bibfnamefont {F.}~\bibnamefont
  {Yan}}, \bibinfo {author} {\bibfnamefont {S.}~\bibnamefont {Gustavsson}},
  \bibinfo {author} {\bibfnamefont {A.}~\bibnamefont {Kamal}}, \bibinfo
  {author} {\bibfnamefont {J.}~\bibnamefont {Birenbaum}}, \bibinfo {author}
  {\bibfnamefont {A.~P.}\ \bibnamefont {Sears}}, \bibinfo {author}
  {\bibfnamefont {D.}~\bibnamefont {Hover}}, \bibinfo {author} {\bibfnamefont
  {T.~J.}\ \bibnamefont {Gudmundsen}}, \bibinfo {author} {\bibfnamefont
  {D.}~\bibnamefont {Rosenberg}}, \bibinfo {author} {\bibfnamefont
  {G.}~\bibnamefont {Samach}}, \bibinfo {author} {\bibfnamefont
  {S.}~\bibnamefont {Weber}}, \bibinfo {author} {\bibfnamefont {J.~L.}\
  \bibnamefont {Yoder}}, \bibinfo {author} {\bibfnamefont {T.~P.}\ \bibnamefont
  {Orlando}}, \bibinfo {author} {\bibfnamefont {J.}~\bibnamefont {Clarke}},
  \bibinfo {author} {\bibfnamefont {A.~J.}\ \bibnamefont {Kerman}},\ and\
  \bibinfo {author} {\bibfnamefont {W.~D.}\ \bibnamefont {Oliver}},\ }\href
  {https://doi.org/10.1038/ncomms12964} {\bibfield  {journal} {\bibinfo
  {journal} {Nature Communications}\ }\textbf {\bibinfo {volume} {7}},\
  \bibinfo {pages} {12964} (\bibinfo {year} {2016})}\BibitemShut {NoStop}%
\bibitem [{\citenamefont {Quintana}\ \emph {et~al.}(2017)\citenamefont
  {Quintana}, \citenamefont {Chen}, \citenamefont {Sank}, \citenamefont
  {Petukhov}, \citenamefont {White}, \citenamefont {Kafri}, \citenamefont
  {Chiaro}, \citenamefont {Megrant}, \citenamefont {Barends}, \citenamefont
  {Campbell}, \citenamefont {Chen}, \citenamefont {Dunsworth}, \citenamefont
  {Fowler}, \citenamefont {Graff}, \citenamefont {Jeffrey}, \citenamefont
  {Kelly}, \citenamefont {Lucero}, \citenamefont {Mutus}, \citenamefont
  {Neeley}, \citenamefont {Neill}, \citenamefont {O'Malley}, \citenamefont
  {Roushan}, \citenamefont {Shabani}, \citenamefont {Smelyanskiy},
  \citenamefont {Vainsencher}, \citenamefont {Wenner}, \citenamefont {Neven},\
  and\ \citenamefont {Martinis}}]{Quintana_2017}%
  \BibitemOpen
  \bibfield  {author} {\bibinfo {author} {\bibfnamefont {C.~M.}\ \bibnamefont
  {Quintana}}, \bibinfo {author} {\bibfnamefont {Y.}~\bibnamefont {Chen}},
  \bibinfo {author} {\bibfnamefont {D.}~\bibnamefont {Sank}}, \bibinfo {author}
  {\bibfnamefont {A.~G.}\ \bibnamefont {Petukhov}}, \bibinfo {author}
  {\bibfnamefont {T.~C.}\ \bibnamefont {White}}, \bibinfo {author}
  {\bibfnamefont {D.}~\bibnamefont {Kafri}}, \bibinfo {author} {\bibfnamefont
  {B.}~\bibnamefont {Chiaro}}, \bibinfo {author} {\bibfnamefont
  {A.}~\bibnamefont {Megrant}}, \bibinfo {author} {\bibfnamefont
  {R.}~\bibnamefont {Barends}}, \bibinfo {author} {\bibfnamefont
  {B.}~\bibnamefont {Campbell}}, \bibinfo {author} {\bibfnamefont
  {Z.}~\bibnamefont {Chen}}, \bibinfo {author} {\bibfnamefont {A.}~\bibnamefont
  {Dunsworth}}, \bibinfo {author} {\bibfnamefont {A.~G.}\ \bibnamefont
  {Fowler}}, \bibinfo {author} {\bibfnamefont {R.}~\bibnamefont {Graff}},
  \bibinfo {author} {\bibfnamefont {E.}~\bibnamefont {Jeffrey}}, \bibinfo
  {author} {\bibfnamefont {J.}~\bibnamefont {Kelly}}, \bibinfo {author}
  {\bibfnamefont {E.}~\bibnamefont {Lucero}}, \bibinfo {author} {\bibfnamefont
  {J.~Y.}\ \bibnamefont {Mutus}}, \bibinfo {author} {\bibfnamefont
  {M.}~\bibnamefont {Neeley}}, \bibinfo {author} {\bibfnamefont
  {C.}~\bibnamefont {Neill}}, \bibinfo {author} {\bibfnamefont {P.~J.~J.}\
  \bibnamefont {O'Malley}}, \bibinfo {author} {\bibfnamefont {P.}~\bibnamefont
  {Roushan}}, \bibinfo {author} {\bibfnamefont {A.}~\bibnamefont {Shabani}},
  \bibinfo {author} {\bibfnamefont {V.~N.}\ \bibnamefont {Smelyanskiy}},
  \bibinfo {author} {\bibfnamefont {A.}~\bibnamefont {Vainsencher}}, \bibinfo
  {author} {\bibfnamefont {J.}~\bibnamefont {Wenner}}, \bibinfo {author}
  {\bibfnamefont {H.}~\bibnamefont {Neven}},\ and\ \bibinfo {author}
  {\bibfnamefont {J.~M.}\ \bibnamefont {Martinis}},\ }\href
  {https://doi.org/10.1103/PhysRevLett.118.057702} {\bibfield  {journal}
  {\bibinfo  {journal} {Phys. Rev. Lett.}\ }\textbf {\bibinfo {volume} {118}},\
  \bibinfo {pages} {057702} (\bibinfo {year} {2017})}\BibitemShut {NoStop}%
\bibitem [{\citenamefont {Valery}\ \emph {et~al.}(2022)\citenamefont {Valery},
  \citenamefont {Chowdhury}, \citenamefont {Jones},\ and\ \citenamefont
  {Didier}}]{Valery_2022}%
  \BibitemOpen
  \bibfield  {author} {\bibinfo {author} {\bibfnamefont {J.~A.}\ \bibnamefont
  {Valery}}, \bibinfo {author} {\bibfnamefont {S.}~\bibnamefont {Chowdhury}},
  \bibinfo {author} {\bibfnamefont {G.}~\bibnamefont {Jones}},\ and\ \bibinfo
  {author} {\bibfnamefont {N.}~\bibnamefont {Didier}},\ }\href
  {https://doi.org/10.1103/PRXQuantum.3.020337} {\bibfield  {journal} {\bibinfo
   {journal} {PRX Quantum}\ }\textbf {\bibinfo {volume} {3}},\ \bibinfo {pages}
  {020337} (\bibinfo {year} {2022})}\BibitemShut {NoStop}%
\bibitem [{\citenamefont {Sarma}\ \emph {et~al.}(2015)\citenamefont {Sarma},
  \citenamefont {Freedman},\ and\ \citenamefont {Nayak}}]{Sarma:2015aa}%
  \BibitemOpen
  \bibfield  {author} {\bibinfo {author} {\bibfnamefont {S.~D.}\ \bibnamefont
  {Sarma}}, \bibinfo {author} {\bibfnamefont {M.}~\bibnamefont {Freedman}},\
  and\ \bibinfo {author} {\bibfnamefont {C.}~\bibnamefont {Nayak}},\ }\href
  {https://doi.org/https://doi.org/10.1038/npjqi.2015.1} {\bibfield  {journal}
  {\bibinfo  {journal} {npj Quantum Information}\ }\textbf {\bibinfo {volume}
  {1}},\ \bibinfo {pages} {15001} (\bibinfo {year} {2015})}\BibitemShut
  {NoStop}%
\bibitem [{\citenamefont {Bansil}\ \emph {et~al.}(2016)\citenamefont {Bansil},
  \citenamefont {Lin},\ and\ \citenamefont {Das}}]{Bansil:2016aa}%
  \BibitemOpen
  \bibfield  {author} {\bibinfo {author} {\bibfnamefont {A.}~\bibnamefont
  {Bansil}}, \bibinfo {author} {\bibfnamefont {H.}~\bibnamefont {Lin}},\ and\
  \bibinfo {author} {\bibfnamefont {T.}~\bibnamefont {Das}},\ }\href
  {https://doi.org/10.1103/RevModPhys.88.021004} {\bibfield  {journal}
  {\bibinfo  {journal} {Rev. Mod. Phys.}\ }\textbf {\bibinfo {volume} {88}},\
  \bibinfo {pages} {021004} (\bibinfo {year} {2016})}\BibitemShut {NoStop}%
\bibitem [{\citenamefont {Hatsugai}(1993)}]{Hatsugai:1993aa}%
  \BibitemOpen
  \bibfield  {author} {\bibinfo {author} {\bibfnamefont {Y.}~\bibnamefont
  {Hatsugai}},\ }\href {https://doi.org/10.1103/PhysRevLett.71.3697} {\bibfield
   {journal} {\bibinfo  {journal} {Phys. Rev. Lett.}\ }\textbf {\bibinfo
  {volume} {71}},\ \bibinfo {pages} {3697} (\bibinfo {year}
  {1993})}\BibitemShut {NoStop}%
\bibitem [{\citenamefont {Kitaev}(2001)}]{Kitaev2001}%
  \BibitemOpen
  \bibfield  {author} {\bibinfo {author} {\bibfnamefont {A.~Y.}\ \bibnamefont
  {Kitaev}},\ }\href
  {https://doi.org/https://doi.org/10.1070/1063-7869/44/10s/s29} {\bibfield
  {journal} {\bibinfo  {journal} {Physics-Uspekhi}\ }\textbf {\bibinfo {volume}
  {44}},\ \bibinfo {pages} {131} (\bibinfo {year} {2001})}\BibitemShut
  {NoStop}%
\bibitem [{\citenamefont {Kane}\ and\ \citenamefont
  {Mele}(2005)}]{Kane2005TopologicalOrder}%
  \BibitemOpen
  \bibfield  {author} {\bibinfo {author} {\bibfnamefont {C.~L.}\ \bibnamefont
  {Kane}}\ and\ \bibinfo {author} {\bibfnamefont {E.~J.}\ \bibnamefont
  {Mele}},\ }\href {https://doi.org/10.1103/PhysRevLett.95.146802} {\bibfield
  {journal} {\bibinfo  {journal} {Phys. Rev. Lett.}\ }\textbf {\bibinfo
  {volume} {95}},\ \bibinfo {pages} {146802} (\bibinfo {year}
  {2005})}\BibitemShut {NoStop}%
\bibitem [{\citenamefont {Bernevig}\ and\ \citenamefont
  {Zhang}(2006)}]{Bernevig:2006aa}%
  \BibitemOpen
  \bibfield  {author} {\bibinfo {author} {\bibfnamefont {B.~A.}\ \bibnamefont
  {Bernevig}}\ and\ \bibinfo {author} {\bibfnamefont {S.-C.}\ \bibnamefont
  {Zhang}},\ }\href {https://doi.org/10.1103/PhysRevLett.96.106802} {\bibfield
  {journal} {\bibinfo  {journal} {Phys. Rev. Lett.}\ }\textbf {\bibinfo
  {volume} {96}},\ \bibinfo {pages} {106802} (\bibinfo {year}
  {2006})}\BibitemShut {NoStop}%
\bibitem [{\citenamefont {Fu}\ \emph {et~al.}(2007)\citenamefont {Fu},
  \citenamefont {Kane},\ and\ \citenamefont {Mele}}]{Fu:2007aa}%
  \BibitemOpen
  \bibfield  {author} {\bibinfo {author} {\bibfnamefont {L.}~\bibnamefont
  {Fu}}, \bibinfo {author} {\bibfnamefont {C.~L.}\ \bibnamefont {Kane}},\ and\
  \bibinfo {author} {\bibfnamefont {E.~J.}\ \bibnamefont {Mele}},\ }\href
  {https://doi.org/10.1103/PhysRevLett.98.106803} {\bibfield  {journal}
  {\bibinfo  {journal} {Phys. Rev. Lett.}\ }\textbf {\bibinfo {volume} {98}},\
  \bibinfo {pages} {106803} (\bibinfo {year} {2007})}\BibitemShut {NoStop}%
\bibitem [{\citenamefont {Murakami}(2007)}]{Murakami:2007aa}%
  \BibitemOpen
  \bibfield  {author} {\bibinfo {author} {\bibfnamefont {S.}~\bibnamefont
  {Murakami}},\ }\href {https://doi.org/10.1088/1367-2630/9/9/356} {\bibfield
  {journal} {\bibinfo  {journal} {New Journal of Physics}\ }\textbf {\bibinfo
  {volume} {9}},\ \bibinfo {pages} {356} (\bibinfo {year} {2007})}\BibitemShut
  {NoStop}%
\bibitem [{\citenamefont {Fu}\ and\ \citenamefont {Kane}(2008)}]{Fu:2008aa}%
  \BibitemOpen
  \bibfield  {author} {\bibinfo {author} {\bibfnamefont {L.}~\bibnamefont
  {Fu}}\ and\ \bibinfo {author} {\bibfnamefont {C.~L.}\ \bibnamefont {Kane}},\
  }\href {https://doi.org/10.1103/PhysRevLett.100.096407} {\bibfield  {journal}
  {\bibinfo  {journal} {Phys. Rev. Lett.}\ }\textbf {\bibinfo {volume} {100}},\
  \bibinfo {pages} {096407} (\bibinfo {year} {2008})}\BibitemShut {NoStop}%
\bibitem [{\citenamefont {Prodan}\ \emph {et~al.}(2010)\citenamefont {Prodan},
  \citenamefont {Hughes},\ and\ \citenamefont {Bernevig}}]{Prodan:2010aa}%
  \BibitemOpen
  \bibfield  {author} {\bibinfo {author} {\bibfnamefont {E.}~\bibnamefont
  {Prodan}}, \bibinfo {author} {\bibfnamefont {T.~L.}\ \bibnamefont {Hughes}},\
  and\ \bibinfo {author} {\bibfnamefont {B.~A.}\ \bibnamefont {Bernevig}},\
  }\href {https://doi.org/10.1103/PhysRevLett.105.115501} {\bibfield  {journal}
  {\bibinfo  {journal} {Phys. Rev. Lett.}\ }\textbf {\bibinfo {volume} {105}},\
  \bibinfo {pages} {115501} (\bibinfo {year} {2010})}\BibitemShut {NoStop}%
\bibitem [{\citenamefont {Hasan}\ and\ \citenamefont
  {Kane}(2010)}]{Hasan:2010aa}%
  \BibitemOpen
  \bibfield  {author} {\bibinfo {author} {\bibfnamefont {M.~Z.}\ \bibnamefont
  {Hasan}}\ and\ \bibinfo {author} {\bibfnamefont {C.~L.}\ \bibnamefont
  {Kane}},\ }\href {https://doi.org/10.1103/RevModPhys.82.3045} {\bibfield
  {journal} {\bibinfo  {journal} {Rev. Mod. Phys.}\ }\textbf {\bibinfo {volume}
  {82}},\ \bibinfo {pages} {3045} (\bibinfo {year} {2010})}\BibitemShut
  {NoStop}%
\bibitem [{\citenamefont {Qi}\ and\ \citenamefont {Zhang}(2011)}]{Qi:2011aa}%
  \BibitemOpen
  \bibfield  {author} {\bibinfo {author} {\bibfnamefont {X.-L.}\ \bibnamefont
  {Qi}}\ and\ \bibinfo {author} {\bibfnamefont {S.-C.}\ \bibnamefont {Zhang}},\
  }\href {https://doi.org/https://doi.org/10.1103/RevModPhys.83.1057}
  {\bibfield  {journal} {\bibinfo  {journal} {Rev. Mod. Phys.}\ }\textbf
  {\bibinfo {volume} {83}},\ \bibinfo {pages} {1057} (\bibinfo {year}
  {2011})}\BibitemShut {NoStop}%
\bibitem [{\citenamefont {Wan}\ \emph {et~al.}(2011)\citenamefont {Wan},
  \citenamefont {Turner}, \citenamefont {Vishwanath},\ and\ \citenamefont
  {Savrasov}}]{Wan:2011aa}%
  \BibitemOpen
  \bibfield  {author} {\bibinfo {author} {\bibfnamefont {X.}~\bibnamefont
  {Wan}}, \bibinfo {author} {\bibfnamefont {A.~M.}\ \bibnamefont {Turner}},
  \bibinfo {author} {\bibfnamefont {A.}~\bibnamefont {Vishwanath}},\ and\
  \bibinfo {author} {\bibfnamefont {S.~Y.}\ \bibnamefont {Savrasov}},\ }\href
  {https://doi.org/https://doi.org/10.1103/PhysRevB.83.205101} {\bibfield
  {journal} {\bibinfo  {journal} {Phys. Rev. B}\ }\textbf {\bibinfo {volume}
  {83}},\ \bibinfo {pages} {205101} (\bibinfo {year} {2011})}\BibitemShut
  {NoStop}%
\bibitem [{\citenamefont {Burkov}\ and\ \citenamefont
  {Balents}(2011)}]{Burkov:2011aa}%
  \BibitemOpen
  \bibfield  {author} {\bibinfo {author} {\bibfnamefont {A.~A.}\ \bibnamefont
  {Burkov}}\ and\ \bibinfo {author} {\bibfnamefont {L.}~\bibnamefont
  {Balents}},\ }\href
  {https://doi.org/https://doi.org/10.1103/PhysRevLett.107.127205} {\bibfield
  {journal} {\bibinfo  {journal} {Phys. Rev. Lett.}\ }\textbf {\bibinfo
  {volume} {107}},\ \bibinfo {pages} {127205} (\bibinfo {year}
  {2011})}\BibitemShut {NoStop}%
\bibitem [{\citenamefont {Meng}\ and\ \citenamefont
  {Balents}(2012)}]{Meng:2012aa}%
  \BibitemOpen
  \bibfield  {author} {\bibinfo {author} {\bibfnamefont {T.}~\bibnamefont
  {Meng}}\ and\ \bibinfo {author} {\bibfnamefont {L.}~\bibnamefont {Balents}},\
  }\href {https://doi.org/10.1103/PhysRevB.86.054504} {\bibfield  {journal}
  {\bibinfo  {journal} {Phys. Rev. B}\ }\textbf {\bibinfo {volume} {86}},\
  \bibinfo {pages} {054504} (\bibinfo {year} {2012})}\BibitemShut {NoStop}%
\bibitem [{\citenamefont {Alicea}(2012)}]{Alicea:2012aa}%
  \BibitemOpen
  \bibfield  {author} {\bibinfo {author} {\bibfnamefont {J.}~\bibnamefont
  {Alicea}},\ }\href
  {https://doi.org/https://doi.org/10.1088/0034-4885/75/7/076501} {\bibfield
  {journal} {\bibinfo  {journal} {Reports on Progress in Physics}\ }\textbf
  {\bibinfo {volume} {75}},\ \bibinfo {pages} {076501} (\bibinfo {year}
  {2012})}\BibitemShut {NoStop}%
\bibitem [{\citenamefont {Hosur}\ and\ \citenamefont
  {Qi}(2013)}]{Hosur:2013aa}%
  \BibitemOpen
  \bibfield  {author} {\bibinfo {author} {\bibfnamefont {P.}~\bibnamefont
  {Hosur}}\ and\ \bibinfo {author} {\bibfnamefont {X.}~\bibnamefont {Qi}},\
  }\bibfield  {booktitle} {\emph {\bibinfo {booktitle} {Topological insulators
  / Isolants topologiques}},\ }\href
  {https://doi.org/https://doi.org/10.1016/j.crhy.2013.10.010} {\bibfield
  {journal} {\bibinfo  {journal} {Comptes Rendus Physique}\ }\textbf {\bibinfo
  {volume} {14}},\ \bibinfo {pages} {857} (\bibinfo {year} {2013})}\BibitemShut
  {NoStop}%
\bibitem [{\citenamefont {Yang}\ \emph {et~al.}(2014)\citenamefont {Yang},
  \citenamefont {Pan},\ and\ \citenamefont {Zhang}}]{Yang:2014aa}%
  \BibitemOpen
  \bibfield  {author} {\bibinfo {author} {\bibfnamefont {S.~A.}\ \bibnamefont
  {Yang}}, \bibinfo {author} {\bibfnamefont {H.}~\bibnamefont {Pan}},\ and\
  \bibinfo {author} {\bibfnamefont {F.}~\bibnamefont {Zhang}},\ }\href
  {https://link.aps.org/doi/10.1103/PhysRevLett.113.046401} {\bibfield
  {journal} {\bibinfo  {journal} {Phys. Rev. Lett.}\ }\textbf {\bibinfo
  {volume} {113}},\ \bibinfo {pages} {046401} (\bibinfo {year}
  {2014})}\BibitemShut {NoStop}%
\bibitem [{\citenamefont {Bednik}\ \emph {et~al.}(2015)\citenamefont {Bednik},
  \citenamefont {Zyuzin},\ and\ \citenamefont {Burkov}}]{Bednik:2015aa}%
  \BibitemOpen
  \bibfield  {author} {\bibinfo {author} {\bibfnamefont {G.}~\bibnamefont
  {Bednik}}, \bibinfo {author} {\bibfnamefont {A.~A.}\ \bibnamefont {Zyuzin}},\
  and\ \bibinfo {author} {\bibfnamefont {A.~A.}\ \bibnamefont {Burkov}},\
  }\href {https://link.aps.org/doi/10.1103/PhysRevB.92.035153} {\bibfield
  {journal} {\bibinfo  {journal} {Phys. Rev. B}\ }\textbf {\bibinfo {volume}
  {92}},\ \bibinfo {pages} {035153} (\bibinfo {year} {2015})}\BibitemShut
  {NoStop}%
\bibitem [{\citenamefont {Lv}\ \emph {et~al.}(2015)\citenamefont {Lv},
  \citenamefont {Weng}, \citenamefont {Fu}, \citenamefont {Wang}, \citenamefont
  {Miao}, \citenamefont {Ma}, \citenamefont {Richard}, \citenamefont {Huang},
  \citenamefont {Zhao}, \citenamefont {Chen}, \citenamefont {Fang},
  \citenamefont {Dai}, \citenamefont {Qian},\ and\ \citenamefont
  {Ding}}]{Lv:2015aa}%
  \BibitemOpen
  \bibfield  {author} {\bibinfo {author} {\bibfnamefont {B.~Q.}\ \bibnamefont
  {Lv}}, \bibinfo {author} {\bibfnamefont {H.~M.}\ \bibnamefont {Weng}},
  \bibinfo {author} {\bibfnamefont {B.~B.}\ \bibnamefont {Fu}}, \bibinfo
  {author} {\bibfnamefont {X.~P.}\ \bibnamefont {Wang}}, \bibinfo {author}
  {\bibfnamefont {H.}~\bibnamefont {Miao}}, \bibinfo {author} {\bibfnamefont
  {J.}~\bibnamefont {Ma}}, \bibinfo {author} {\bibfnamefont {P.}~\bibnamefont
  {Richard}}, \bibinfo {author} {\bibfnamefont {X.~C.}\ \bibnamefont {Huang}},
  \bibinfo {author} {\bibfnamefont {L.~X.}\ \bibnamefont {Zhao}}, \bibinfo
  {author} {\bibfnamefont {G.~F.}\ \bibnamefont {Chen}}, \bibinfo {author}
  {\bibfnamefont {Z.}~\bibnamefont {Fang}}, \bibinfo {author} {\bibfnamefont
  {X.}~\bibnamefont {Dai}}, \bibinfo {author} {\bibfnamefont {T.}~\bibnamefont
  {Qian}},\ and\ \bibinfo {author} {\bibfnamefont {H.}~\bibnamefont {Ding}},\
  }\href {https://doi.org/https://doi.org/10.1103/PhysRevX.5.031013} {\bibfield
   {journal} {\bibinfo  {journal} {Phys. Rev. X}\ }\textbf {\bibinfo {volume}
  {5}},\ \bibinfo {pages} {031013} (\bibinfo {year} {2015})}\BibitemShut
  {NoStop}%
\bibitem [{\citenamefont {Chiu}\ \emph {et~al.}(2016)\citenamefont {Chiu},
  \citenamefont {Teo}, \citenamefont {Schnyder},\ and\ \citenamefont
  {Ryu}}]{Chiu:2016aa}%
  \BibitemOpen
  \bibfield  {author} {\bibinfo {author} {\bibfnamefont {C.-K.}\ \bibnamefont
  {Chiu}}, \bibinfo {author} {\bibfnamefont {J.~C.~Y.}\ \bibnamefont {Teo}},
  \bibinfo {author} {\bibfnamefont {A.~P.}\ \bibnamefont {Schnyder}},\ and\
  \bibinfo {author} {\bibfnamefont {S.}~\bibnamefont {Ryu}},\ }\href
  {https://link.aps.org/doi/10.1103/RevModPhys.88.035005} {\bibfield  {journal}
  {\bibinfo  {journal} {Rev. Mod. Phys.}\ }\textbf {\bibinfo {volume} {88}},\
  \bibinfo {pages} {035005} (\bibinfo {year} {2016})}\BibitemShut {NoStop}%
\bibitem [{\citenamefont {Sato}\ and\ \citenamefont
  {Ando}(2017)}]{Sato:2017aa}%
  \BibitemOpen
  \bibfield  {author} {\bibinfo {author} {\bibfnamefont {M.}~\bibnamefont
  {Sato}}\ and\ \bibinfo {author} {\bibfnamefont {Y.}~\bibnamefont {Ando}},\
  }\href {https://doi.org/https://doi.org/10.1088/1361-6633/aa6ac7} {\bibfield
  {journal} {\bibinfo  {journal} {Reports on Progress in Physics}\ }\textbf
  {\bibinfo {volume} {80}},\ \bibinfo {pages} {076501} (\bibinfo {year}
  {2017})}\BibitemShut {NoStop}%
\bibitem [{\citenamefont {Bernevig}\ \emph {et~al.}(2018)\citenamefont
  {Bernevig}, \citenamefont {Weng}, \citenamefont {Fang},\ and\ \citenamefont
  {Dai}}]{Bernevig:2018aa}%
  \BibitemOpen
  \bibfield  {author} {\bibinfo {author} {\bibfnamefont {A.}~\bibnamefont
  {Bernevig}}, \bibinfo {author} {\bibfnamefont {H.}~\bibnamefont {Weng}},
  \bibinfo {author} {\bibfnamefont {Z.}~\bibnamefont {Fang}},\ and\ \bibinfo
  {author} {\bibfnamefont {X.}~\bibnamefont {Dai}},\ }\href
  {https://doi.org/https://doi.org/10.7566/JPSJ.87.041001} {\bibfield
  {journal} {\bibinfo  {journal} {J. Phys. Soc. Jpn.}\ }\textbf {\bibinfo
  {volume} {87}},\ \bibinfo {pages} {041001} (\bibinfo {year}
  {2018})}\BibitemShut {NoStop}%
\bibitem [{\citenamefont {Armitage}\ \emph {et~al.}(2018)\citenamefont
  {Armitage}, \citenamefont {Mele},\ and\ \citenamefont
  {Vishwanath}}]{Armitage:2018aa}%
  \BibitemOpen
  \bibfield  {author} {\bibinfo {author} {\bibfnamefont {N.~P.}\ \bibnamefont
  {Armitage}}, \bibinfo {author} {\bibfnamefont {E.~J.}\ \bibnamefont {Mele}},\
  and\ \bibinfo {author} {\bibfnamefont {A.}~\bibnamefont {Vishwanath}},\
  }\href {https://doi.org/https://doi.org/10.1103/RevModPhys.90.015001}
  {\bibfield  {journal} {\bibinfo  {journal} {Rev. Mod. Phys.}\ }\textbf
  {\bibinfo {volume} {90}},\ \bibinfo {pages} {015001} (\bibinfo {year}
  {2018})}\BibitemShut {NoStop}%
\bibitem [{\citenamefont {Lutchyn}\ \emph {et~al.}(2018)\citenamefont
  {Lutchyn}, \citenamefont {Bakkers}, \citenamefont {Kouwenhoven},
  \citenamefont {Krogstrup}, \citenamefont {Marcus},\ and\ \citenamefont
  {Oreg}}]{Lutchyn:2018aa}%
  \BibitemOpen
  \bibfield  {author} {\bibinfo {author} {\bibfnamefont {R.~M.}\ \bibnamefont
  {Lutchyn}}, \bibinfo {author} {\bibfnamefont {E.~P. A.~M.}\ \bibnamefont
  {Bakkers}}, \bibinfo {author} {\bibfnamefont {L.~P.}\ \bibnamefont
  {Kouwenhoven}}, \bibinfo {author} {\bibfnamefont {P.}~\bibnamefont
  {Krogstrup}}, \bibinfo {author} {\bibfnamefont {C.~M.}\ \bibnamefont
  {Marcus}},\ and\ \bibinfo {author} {\bibfnamefont {Y.}~\bibnamefont {Oreg}},\
  }\href {https://doi.org/https://doi.org/10.1038/s41578-018-0003-1} {\bibfield
   {journal} {\bibinfo  {journal} {Nature Reviews Materials}\ }\textbf
  {\bibinfo {volume} {3}},\ \bibinfo {pages} {52} (\bibinfo {year}
  {2018})}\BibitemShut {NoStop}%
\bibitem [{\citenamefont {Peralta~Gavensky}\ \emph {et~al.}(2019)\citenamefont
  {Peralta~Gavensky}, \citenamefont {Usaj},\ and\ \citenamefont
  {Balseiro}}]{Peralta-Gavensky:2019aa}%
  \BibitemOpen
  \bibfield  {author} {\bibinfo {author} {\bibfnamefont {L.}~\bibnamefont
  {Peralta~Gavensky}}, \bibinfo {author} {\bibfnamefont {G.}~\bibnamefont
  {Usaj}},\ and\ \bibinfo {author} {\bibfnamefont {C.~A.}\ \bibnamefont
  {Balseiro}},\ }\href
  {https://doi.org/https://doi.org/10.1103/PhysRevB.100.014514} {\bibfield
  {journal} {\bibinfo  {journal} {Phys. Rev. B}\ }\textbf {\bibinfo {volume}
  {100}},\ \bibinfo {pages} {014514} (\bibinfo {year} {2019})}\BibitemShut
  {NoStop}%
\bibitem [{\citenamefont {Sakurai}\ \emph {et~al.}(2020)\citenamefont
  {Sakurai}, \citenamefont {Mercaldo}, \citenamefont {Kobayashi}, \citenamefont
  {Yamakage}, \citenamefont {Ikegaya}, \citenamefont {Habe}, \citenamefont
  {Kotetes}, \citenamefont {Cuoco},\ and\ \citenamefont
  {Asano}}]{Sakurai:2020aa}%
  \BibitemOpen
  \bibfield  {author} {\bibinfo {author} {\bibfnamefont {K.}~\bibnamefont
  {Sakurai}}, \bibinfo {author} {\bibfnamefont {M.~T.}\ \bibnamefont
  {Mercaldo}}, \bibinfo {author} {\bibfnamefont {S.}~\bibnamefont {Kobayashi}},
  \bibinfo {author} {\bibfnamefont {A.}~\bibnamefont {Yamakage}}, \bibinfo
  {author} {\bibfnamefont {S.}~\bibnamefont {Ikegaya}}, \bibinfo {author}
  {\bibfnamefont {T.}~\bibnamefont {Habe}}, \bibinfo {author} {\bibfnamefont
  {P.}~\bibnamefont {Kotetes}}, \bibinfo {author} {\bibfnamefont
  {M.}~\bibnamefont {Cuoco}},\ and\ \bibinfo {author} {\bibfnamefont
  {Y.}~\bibnamefont {Asano}},\ }\href
  {https://doi.org/https://doi.org/10.1103/PhysRevB.101.174506} {\bibfield
  {journal} {\bibinfo  {journal} {Phys. Rev. B}\ }\textbf {\bibinfo {volume}
  {101}},\ \bibinfo {pages} {174506} (\bibinfo {year} {2020})}\BibitemShut
  {NoStop}%
\bibitem [{\citenamefont {Rui}\ \emph {et~al.}(2021)\citenamefont {Rui},
  \citenamefont {Zhang}, \citenamefont {Hirschmann}, \citenamefont {Zheng},
  \citenamefont {Schnyder}, \citenamefont {Trauzettel},\ and\ \citenamefont
  {Wang}}]{rui2020higherorder}%
  \BibitemOpen
  \bibfield  {author} {\bibinfo {author} {\bibfnamefont {W.~B.}\ \bibnamefont
  {Rui}}, \bibinfo {author} {\bibfnamefont {S.-B.}\ \bibnamefont {Zhang}},
  \bibinfo {author} {\bibfnamefont {M.~M.}\ \bibnamefont {Hirschmann}},
  \bibinfo {author} {\bibfnamefont {Z.}~\bibnamefont {Zheng}}, \bibinfo
  {author} {\bibfnamefont {A.~P.}\ \bibnamefont {Schnyder}}, \bibinfo {author}
  {\bibfnamefont {B.}~\bibnamefont {Trauzettel}},\ and\ \bibinfo {author}
  {\bibfnamefont {Z.~D.}\ \bibnamefont {Wang}},\ }\href
  {https://doi.org/10.1103/PhysRevB.103.184510} {\bibfield  {journal} {\bibinfo
   {journal} {Phys. Rev. B}\ }\textbf {\bibinfo {volume} {103}},\ \bibinfo
  {pages} {184510} (\bibinfo {year} {2021})}\BibitemShut {NoStop}%
\bibitem [{\citenamefont {Leone}\ \emph {et~al.}(2008)\citenamefont {Leone},
  \citenamefont {L{\'e}vy},\ and\ \citenamefont {Lafarge}}]{Leone2008}%
  \BibitemOpen
  \bibfield  {author} {\bibinfo {author} {\bibfnamefont {R.}~\bibnamefont
  {Leone}}, \bibinfo {author} {\bibfnamefont {L.~P.}\ \bibnamefont
  {L{\'e}vy}},\ and\ \bibinfo {author} {\bibfnamefont {P.}~\bibnamefont
  {Lafarge}},\ }\href
  {https://doi.org/https://doi.org/10.1103/PhysRevLett.100.117001} {\bibfield
  {journal} {\bibinfo  {journal} {Phys. Rev. Lett.}\ }\textbf {\bibinfo
  {volume} {100}},\ \bibinfo {pages} {117001} (\bibinfo {year}
  {2008})}\BibitemShut {NoStop}%
\bibitem [{\citenamefont {Leone}\ and\ \citenamefont
  {Monjou}(2013)}]{Leone_2013}%
  \BibitemOpen
  \bibfield  {author} {\bibinfo {author} {\bibnamefont {Leone}}\ and\ \bibinfo
  {author} {\bibnamefont {Monjou}},\ }\href
  {https://doi.org/https://doi.org/10.5488/cmp.16.33801} {\bibfield  {journal}
  {\bibinfo  {journal} {Condensed Matter Physics}\ }\textbf {\bibinfo {volume}
  {16}},\ \bibinfo {pages} {33801} (\bibinfo {year} {2013})}\BibitemShut
  {NoStop}%
\bibitem [{\citenamefont {Yokoyama}\ and\ \citenamefont
  {Nazarov}(2015)}]{Yokoyama2015TopolABSmultitJJ}%
  \BibitemOpen
  \bibfield  {author} {\bibinfo {author} {\bibfnamefont {T.}~\bibnamefont
  {Yokoyama}}\ and\ \bibinfo {author} {\bibfnamefont {Y.~V.}\ \bibnamefont
  {Nazarov}},\ }\href
  {https://doi.org/https://doi.org/10.1103/PhysRevB.92.155437} {\bibfield
  {journal} {\bibinfo  {journal} {Phys. Rev. B}\ }\textbf {\bibinfo {volume}
  {92}},\ \bibinfo {pages} {155437} (\bibinfo {year} {2015})}\BibitemShut
  {NoStop}%
\bibitem [{\citenamefont {Riwar}\ \emph {et~al.}(2016)\citenamefont {Riwar},
  \citenamefont {Houzet}, \citenamefont {Meyer},\ and\ \citenamefont
  {Nazarov}}]{Riwar2016}%
  \BibitemOpen
  \bibfield  {author} {\bibinfo {author} {\bibfnamefont {R.-P.}\ \bibnamefont
  {Riwar}}, \bibinfo {author} {\bibfnamefont {M.}~\bibnamefont {Houzet}},
  \bibinfo {author} {\bibfnamefont {J.~S.}\ \bibnamefont {Meyer}},\ and\
  \bibinfo {author} {\bibfnamefont {Y.~V.}\ \bibnamefont {Nazarov}},\ }\href
  {https://doi.org/https://doi.org/10.1038/ncomms11167} {\bibfield  {journal}
  {\bibinfo  {journal} {Nature Communications}\ }\textbf {\bibinfo {volume}
  {7}},\ \bibinfo {pages} {11167} (\bibinfo {year} {2016})}\BibitemShut
  {NoStop}%
\bibitem [{\citenamefont {Strambini}\ \emph {et~al.}(2016)\citenamefont
  {Strambini}, \citenamefont {D'Ambrosio}, \citenamefont {Vischi},
  \citenamefont {Bergeret}, \citenamefont {Nazarov},\ and\ \citenamefont
  {Giazotto}}]{Strambini_2016}%
  \BibitemOpen
  \bibfield  {author} {\bibinfo {author} {\bibfnamefont {E.}~\bibnamefont
  {Strambini}}, \bibinfo {author} {\bibfnamefont {S.}~\bibnamefont
  {D'Ambrosio}}, \bibinfo {author} {\bibfnamefont {F.}~\bibnamefont {Vischi}},
  \bibinfo {author} {\bibfnamefont {F.~S.}\ \bibnamefont {Bergeret}}, \bibinfo
  {author} {\bibfnamefont {Y.~V.}\ \bibnamefont {Nazarov}},\ and\ \bibinfo
  {author} {\bibfnamefont {F.}~\bibnamefont {Giazotto}},\ }\href
  {https://doi.org/https://doi.org/10.1038/nnano.2016.157} {\bibfield
  {journal} {\bibinfo  {journal} {Nature Nanotechnology}\ }\textbf {\bibinfo
  {volume} {11}},\ \bibinfo {pages} {1055} (\bibinfo {year}
  {2016})}\BibitemShut {NoStop}%
\bibitem [{\citenamefont {Eriksson}\ \emph {et~al.}(2017)\citenamefont
  {Eriksson}, \citenamefont {Riwar}, \citenamefont {Houzet}, \citenamefont
  {Meyer},\ and\ \citenamefont {Nazarov}}]{Eriksson2017}%
  \BibitemOpen
  \bibfield  {author} {\bibinfo {author} {\bibfnamefont {E.}~\bibnamefont
  {Eriksson}}, \bibinfo {author} {\bibfnamefont {R.-P.}\ \bibnamefont {Riwar}},
  \bibinfo {author} {\bibfnamefont {M.}~\bibnamefont {Houzet}}, \bibinfo
  {author} {\bibfnamefont {J.~S.}\ \bibnamefont {Meyer}},\ and\ \bibinfo
  {author} {\bibfnamefont {Y.~V.}\ \bibnamefont {Nazarov}},\ }\href
  {https://doi.org/https://doi.org/10.1103/PhysRevB.95.075417} {\bibfield
  {journal} {\bibinfo  {journal} {Phys. Rev. B}\ }\textbf {\bibinfo {volume}
  {95}},\ \bibinfo {pages} {075417} (\bibinfo {year} {2017})}\BibitemShut
  {NoStop}%
\bibitem [{\citenamefont {Meyer}\ and\ \citenamefont
  {Houzet}(2017)}]{Meyer:2017aa}%
  \BibitemOpen
  \bibfield  {author} {\bibinfo {author} {\bibfnamefont {J.~S.}\ \bibnamefont
  {Meyer}}\ and\ \bibinfo {author} {\bibfnamefont {M.}~\bibnamefont {Houzet}},\
  }\href {https://doi.org/https://doi.org/10.1103/PhysRevLett.119.136807}
  {\bibfield  {journal} {\bibinfo  {journal} {Phys. Rev. Lett.}\ }\textbf
  {\bibinfo {volume} {119}},\ \bibinfo {pages} {136807} (\bibinfo {year}
  {2017})}\BibitemShut {NoStop}%
\bibitem [{\citenamefont {Deb}\ \emph {et~al.}(2018)\citenamefont {Deb},
  \citenamefont {Sengupta},\ and\ \citenamefont {Sen}}]{Deb:2018aa}%
  \BibitemOpen
  \bibfield  {author} {\bibinfo {author} {\bibfnamefont {O.}~\bibnamefont
  {Deb}}, \bibinfo {author} {\bibfnamefont {K.}~\bibnamefont {Sengupta}},\ and\
  \bibinfo {author} {\bibfnamefont {D.}~\bibnamefont {Sen}},\ }\href
  {https://doi.org/10.1103/PhysRevB.97.174518} {\bibfield  {journal} {\bibinfo
  {journal} {Phys. Rev. B}\ }\textbf {\bibinfo {volume} {97}},\ \bibinfo
  {pages} {174518} (\bibinfo {year} {2018})}\BibitemShut {NoStop}%
\bibitem [{\citenamefont {Riwar}(2019)}]{Riwar_2019b}%
  \BibitemOpen
  \bibfield  {author} {\bibinfo {author} {\bibfnamefont {R.-P.}\ \bibnamefont
  {Riwar}},\ }\href
  {https://doi.org/https://doi.org/10.1103/PhysRevB.100.245416} {\bibfield
  {journal} {\bibinfo  {journal} {Phys. Rev. B}\ }\textbf {\bibinfo {volume}
  {100}},\ \bibinfo {pages} {245416} (\bibinfo {year} {2019})}\BibitemShut
  {NoStop}%
\bibitem [{\citenamefont {Repin}\ \emph {et~al.}(2019)\citenamefont {Repin},
  \citenamefont {Chen},\ and\ \citenamefont {Nazarov}}]{Repin:2019aa}%
  \BibitemOpen
  \bibfield  {author} {\bibinfo {author} {\bibfnamefont {E.~V.}\ \bibnamefont
  {Repin}}, \bibinfo {author} {\bibfnamefont {Y.}~\bibnamefont {Chen}},\ and\
  \bibinfo {author} {\bibfnamefont {Y.~V.}\ \bibnamefont {Nazarov}},\ }\href
  {https://doi.org/https://doi.org/10.1103/PhysRevB.99.165414} {\bibfield
  {journal} {\bibinfo  {journal} {Phys. Rev. B}\ }\textbf {\bibinfo {volume}
  {99}},\ \bibinfo {pages} {165414} (\bibinfo {year} {2019})}\BibitemShut
  {NoStop}%
\bibitem [{\citenamefont {Repin}\ and\ \citenamefont
  {Nazarov}(2022)}]{Repin_2020}%
  \BibitemOpen
  \bibfield  {author} {\bibinfo {author} {\bibfnamefont {E.~V.}\ \bibnamefont
  {Repin}}\ and\ \bibinfo {author} {\bibfnamefont {Y.~V.}\ \bibnamefont
  {Nazarov}},\ }\href
  {https://doi.org/https://doi.org/10.1103/PhysRevB.105.L041405} {\bibfield
  {journal} {\bibinfo  {journal} {Phys. Rev. B}\ }\textbf {\bibinfo {volume}
  {105}},\ \bibinfo {pages} {L041405} (\bibinfo {year} {2022})}\BibitemShut
  {NoStop}%
\bibitem [{\citenamefont {Fatemi}\ \emph {et~al.}(2021)\citenamefont {Fatemi},
  \citenamefont {Akhmerov},\ and\ \citenamefont {Bretheau}}]{Fatemi_2020}%
  \BibitemOpen
  \bibfield  {author} {\bibinfo {author} {\bibfnamefont {V.}~\bibnamefont
  {Fatemi}}, \bibinfo {author} {\bibfnamefont {A.~R.}\ \bibnamefont
  {Akhmerov}},\ and\ \bibinfo {author} {\bibfnamefont {L.}~\bibnamefont
  {Bretheau}},\ }\href
  {https://doi.org/https://doi.org/10.1103/PhysRevResearch.3.013288} {\bibfield
   {journal} {\bibinfo  {journal} {Phys. Rev. Research}\ }\textbf {\bibinfo
  {volume} {3}},\ \bibinfo {pages} {013288} (\bibinfo {year}
  {2021})}\BibitemShut {NoStop}%
\bibitem [{\citenamefont {Peyruchat}\ \emph {et~al.}(2021)\citenamefont
  {Peyruchat}, \citenamefont {Griesmar}, \citenamefont {Pillet},\ and\
  \citenamefont {\c{C}. \"{O}.~Girit}}]{Peyruchat_2020}%
  \BibitemOpen
  \bibfield  {author} {\bibinfo {author} {\bibfnamefont {L.}~\bibnamefont
  {Peyruchat}}, \bibinfo {author} {\bibfnamefont {J.}~\bibnamefont {Griesmar}},
  \bibinfo {author} {\bibfnamefont {J.~D.}\ \bibnamefont {Pillet}},\ and\
  \bibinfo {author} {\bibnamefont {\c{C}. \"{O}.~Girit}},\ }\href
  {https://doi.org/https://doi.org/10.1103/PhysRevResearch.3.013289} {\bibfield
   {journal} {\bibinfo  {journal} {Phys. Rev. Research}\ }\textbf {\bibinfo
  {volume} {3}},\ \bibinfo {pages} {013289} (\bibinfo {year}
  {2021})}\BibitemShut {NoStop}%
\bibitem [{\citenamefont {Klees}\ \emph {et~al.}(2020)\citenamefont {Klees},
  \citenamefont {Rastelli}, \citenamefont {Cuevas},\ and\ \citenamefont
  {Belzig}}]{Klees:2020aa}%
  \BibitemOpen
  \bibfield  {author} {\bibinfo {author} {\bibfnamefont {R.~L.}\ \bibnamefont
  {Klees}}, \bibinfo {author} {\bibfnamefont {G.}~\bibnamefont {Rastelli}},
  \bibinfo {author} {\bibfnamefont {J.~C.}\ \bibnamefont {Cuevas}},\ and\
  \bibinfo {author} {\bibfnamefont {W.}~\bibnamefont {Belzig}},\ }\href
  {https://doi.org/https://doi.org/10.1103/PhysRevLett.124.197002} {\bibfield
  {journal} {\bibinfo  {journal} {Phys. Rev. Lett.}\ }\textbf {\bibinfo
  {volume} {124}},\ \bibinfo {pages} {197002} (\bibinfo {year}
  {2020})}\BibitemShut {NoStop}%
\bibitem [{\citenamefont {Klees}\ \emph {et~al.}(2021)\citenamefont {Klees},
  \citenamefont {Cuevas}, \citenamefont {Belzig},\ and\ \citenamefont
  {Rastelli}}]{Klees_2021}%
  \BibitemOpen
  \bibfield  {author} {\bibinfo {author} {\bibfnamefont {R.~L.}\ \bibnamefont
  {Klees}}, \bibinfo {author} {\bibfnamefont {J.~C.}\ \bibnamefont {Cuevas}},
  \bibinfo {author} {\bibfnamefont {W.}~\bibnamefont {Belzig}},\ and\ \bibinfo
  {author} {\bibfnamefont {G.}~\bibnamefont {Rastelli}},\ }\href
  {https://doi.org/https://doi.org/10.1103/PhysRevB.103.014516} {\bibfield
  {journal} {\bibinfo  {journal} {Phys. Rev. B}\ }\textbf {\bibinfo {volume}
  {103}},\ \bibinfo {pages} {014516} (\bibinfo {year} {2021})}\BibitemShut
  {NoStop}%
\bibitem [{\citenamefont {Weisbrich}\ \emph
  {et~al.}(2021{\natexlab{a}})\citenamefont {Weisbrich}, \citenamefont {Klees},
  \citenamefont {Rastelli},\ and\ \citenamefont {Belzig}}]{Weisbrich_2021}%
  \BibitemOpen
  \bibfield  {author} {\bibinfo {author} {\bibfnamefont {H.}~\bibnamefont
  {Weisbrich}}, \bibinfo {author} {\bibfnamefont {R.~L.}\ \bibnamefont
  {Klees}}, \bibinfo {author} {\bibfnamefont {G.}~\bibnamefont {Rastelli}},\
  and\ \bibinfo {author} {\bibfnamefont {W.}~\bibnamefont {Belzig}},\ }\href
  {https://doi.org/https://doi.org/10.1103/PRXQuantum.2.010310} {\bibfield
  {journal} {\bibinfo  {journal} {PRX Quantum}\ }\textbf {\bibinfo {volume}
  {2}},\ \bibinfo {pages} {010310} (\bibinfo {year}
  {2021}{\natexlab{a}})}\BibitemShut {NoStop}%
\bibitem [{\citenamefont {Weisbrich}\ \emph
  {et~al.}(2021{\natexlab{b}})\citenamefont {Weisbrich}, \citenamefont
  {Bestler},\ and\ \citenamefont {Belzig}}]{Weisbrich_2021_monopoles}%
  \BibitemOpen
  \bibfield  {author} {\bibinfo {author} {\bibfnamefont {H.}~\bibnamefont
  {Weisbrich}}, \bibinfo {author} {\bibfnamefont {M.}~\bibnamefont {Bestler}},\
  and\ \bibinfo {author} {\bibfnamefont {W.}~\bibnamefont {Belzig}},\ }\href
  {https://doi.org/10.22331/q-2021-12-07-601} {\bibfield  {journal} {\bibinfo
  {journal} {Quantum}\ }\textbf {\bibinfo {volume} {5}},\ \bibinfo {pages}
  {601} (\bibinfo {year} {2021}{\natexlab{b}})}\BibitemShut {NoStop}%
\bibitem [{\citenamefont {Chirolli}\ and\ \citenamefont
  {Moore}(2021)}]{Chirolli_2021}%
  \BibitemOpen
  \bibfield  {author} {\bibinfo {author} {\bibfnamefont {L.}~\bibnamefont
  {Chirolli}}\ and\ \bibinfo {author} {\bibfnamefont {J.~E.}\ \bibnamefont
  {Moore}},\ }\href
  {https://doi.org/https://doi.org/10.1103/PhysRevLett.126.187701} {\bibfield
  {journal} {\bibinfo  {journal} {Phys. Rev. Lett.}\ }\textbf {\bibinfo
  {volume} {126}},\ \bibinfo {pages} {187701} (\bibinfo {year}
  {2021})}\BibitemShut {NoStop}%
\bibitem [{\citenamefont {Herrig}\ and\ \citenamefont
  {Riwar}(2022)}]{Herrig_2022}%
  \BibitemOpen
  \bibfield  {author} {\bibinfo {author} {\bibfnamefont {T.}~\bibnamefont
  {Herrig}}\ and\ \bibinfo {author} {\bibfnamefont {R.-P.}\ \bibnamefont
  {Riwar}},\ }\href
  {https://doi.org/https://doi.org/10.1103/PhysRevResearch.4.013038} {\bibfield
   {journal} {\bibinfo  {journal} {Phys. Rev. Research}\ }\textbf {\bibinfo
  {volume} {4}},\ \bibinfo {pages} {013038} (\bibinfo {year}
  {2022})}\BibitemShut {NoStop}%
\bibitem [{\citenamefont {Melo}\ \emph {et~al.}(2022)\citenamefont {Melo},
  \citenamefont {Fatemi},\ and\ \citenamefont {Akhmerov}}]{Melo2022}%
  \BibitemOpen
  \bibfield  {author} {\bibinfo {author} {\bibfnamefont {A.}~\bibnamefont
  {Melo}}, \bibinfo {author} {\bibfnamefont {V.}~\bibnamefont {Fatemi}},\ and\
  \bibinfo {author} {\bibfnamefont {A.~R.}\ \bibnamefont {Akhmerov}},\ }\href
  {https://doi.org/10.21468/SciPostPhys.12.1.017} {\bibfield  {journal}
  {\bibinfo  {journal} {SciPost Phys.}\ }\textbf {\bibinfo {volume} {12}},\
  \bibinfo {pages} {017} (\bibinfo {year} {2022})}\BibitemShut {NoStop}%
\bibitem [{\citenamefont {Javed}\ \emph {et~al.}(2023)\citenamefont {Javed},
  \citenamefont {Schwibbert},\ and\ \citenamefont {Riwar}}]{Javed_2023}%
  \BibitemOpen
  \bibfield  {author} {\bibinfo {author} {\bibfnamefont {M.~A.}\ \bibnamefont
  {Javed}}, \bibinfo {author} {\bibfnamefont {J.}~\bibnamefont {Schwibbert}},\
  and\ \bibinfo {author} {\bibfnamefont {R.-P.}\ \bibnamefont {Riwar}},\ }\href
  {https://doi.org/10.1103/PhysRevB.107.035408} {\bibfield  {journal} {\bibinfo
   {journal} {Phys. Rev. B}\ }\textbf {\bibinfo {volume} {107}},\ \bibinfo
  {pages} {035408} (\bibinfo {year} {2023})}\BibitemShut {NoStop}%
\bibitem [{\citenamefont {Herrig}\ \emph {et~al.}(2022)\citenamefont {Herrig},
  \citenamefont {Pixley}, \citenamefont {König},\ and\ \citenamefont
  {Riwar}}]{herrig2022quasiperiodic}%
  \BibitemOpen
  \bibfield  {author} {\bibinfo {author} {\bibfnamefont {T.}~\bibnamefont
  {Herrig}}, \bibinfo {author} {\bibfnamefont {J.~H.}\ \bibnamefont {Pixley}},
  \bibinfo {author} {\bibfnamefont {E.~J.}\ \bibnamefont {König}},\ and\
  \bibinfo {author} {\bibfnamefont {R.-P.}\ \bibnamefont {Riwar}},\ }\href@noop
  {} {\bibinfo {title} {Quasiperiodic circuit quantum electrodynamics}}
  (\bibinfo {year} {2022}),\ \Eprint {https://arxiv.org/abs/2212.12382}
  {arXiv:2212.12382 [cond-mat.mes-hall]} \BibitemShut {NoStop}%
\bibitem [{\citenamefont {Geim}\ and\ \citenamefont
  {Grigorieva}(2013)}]{GeimGrigorieva2913}%
  \BibitemOpen
  \bibfield  {author} {\bibinfo {author} {\bibfnamefont {A.~K.}\ \bibnamefont
  {Geim}}\ and\ \bibinfo {author} {\bibfnamefont {I.~V.}\ \bibnamefont
  {Grigorieva}},\ }\href {https://doi.org/https://doi.org/10.1038/nature12385}
  {\bibfield  {journal} {\bibinfo  {journal} {Nature}\ }\textbf {\bibinfo
  {volume} {499}},\ \bibinfo {pages} {419} (\bibinfo {year}
  {2013})}\BibitemShut {NoStop}%
\bibitem [{\citenamefont {Cao}\ \emph {et~al.}(2018)\citenamefont {Cao},
  \citenamefont {Fatemi}, \citenamefont {Fang}, \citenamefont {Watanabe},
  \citenamefont {Taniguchi}, \citenamefont {Kaxiras},\ and\ \citenamefont
  {Jarillo-Herrero}}]{CaoJarillo}%
  \BibitemOpen
  \bibfield  {author} {\bibinfo {author} {\bibfnamefont {Y.}~\bibnamefont
  {Cao}}, \bibinfo {author} {\bibfnamefont {V.}~\bibnamefont {Fatemi}},
  \bibinfo {author} {\bibfnamefont {S.}~\bibnamefont {Fang}}, \bibinfo {author}
  {\bibfnamefont {K.}~\bibnamefont {Watanabe}}, \bibinfo {author}
  {\bibfnamefont {T.}~\bibnamefont {Taniguchi}}, \bibinfo {author}
  {\bibfnamefont {E.}~\bibnamefont {Kaxiras}},\ and\ \bibinfo {author}
  {\bibfnamefont {P.}~\bibnamefont {Jarillo-Herrero}},\ }\href
  {https://doi.org/https://doi.org/10.1038/nature26160} {\bibfield  {journal}
  {\bibinfo  {journal} {Nature}\ }\textbf {\bibinfo {volume} {556}},\ \bibinfo
  {pages} {43} (\bibinfo {year} {2018})}\BibitemShut {NoStop}%
\bibitem [{\citenamefont {Gonz\'alez-Tudela}\ and\ \citenamefont
  {Cirac}(2019)}]{GonzalezCirac2019}%
  \BibitemOpen
  \bibfield  {author} {\bibinfo {author} {\bibfnamefont {A.}~\bibnamefont
  {Gonz\'alez-Tudela}}\ and\ \bibinfo {author} {\bibfnamefont {J.~I.}\
  \bibnamefont {Cirac}},\ }\href
  {https://doi.org/https://doi.org/10.1103/PhysRevA.100.053604} {\bibfield
  {journal} {\bibinfo  {journal} {Phys. Rev. A}\ }\textbf {\bibinfo {volume}
  {100}},\ \bibinfo {pages} {053604} (\bibinfo {year} {2019})}\BibitemShut
  {NoStop}%
\bibitem [{\citenamefont {Fu}\ \emph {et~al.}(2020)\citenamefont {Fu},
  \citenamefont {K{\"o}nig}, \citenamefont {Wilson}, \citenamefont {Chou},\
  and\ \citenamefont {Pixley}}]{FuPixley2020}%
  \BibitemOpen
  \bibfield  {author} {\bibinfo {author} {\bibfnamefont {Y.}~\bibnamefont
  {Fu}}, \bibinfo {author} {\bibfnamefont {E.~J.}\ \bibnamefont {K{\"o}nig}},
  \bibinfo {author} {\bibfnamefont {J.~H.}\ \bibnamefont {Wilson}}, \bibinfo
  {author} {\bibfnamefont {Y.-Z.}\ \bibnamefont {Chou}},\ and\ \bibinfo
  {author} {\bibfnamefont {J.~H.}\ \bibnamefont {Pixley}},\ }\href
  {https://doi.org/https://doi.org/10.1038/s41535-020-00271-9} {\bibfield
  {journal} {\bibinfo  {journal} {npj Quantum Materials}\ }\textbf {\bibinfo
  {volume} {5}},\ \bibinfo {pages} {71} (\bibinfo {year} {2020})}\BibitemShut
  {NoStop}%
\bibitem [{\citenamefont {Salamon}\ \emph {et~al.}(2020)\citenamefont
  {Salamon}, \citenamefont {Celi}, \citenamefont {Chhajlany}, \citenamefont
  {Fr\'erot}, \citenamefont {Lewenstein}, \citenamefont {Tarruell},\ and\
  \citenamefont {Rakshit}}]{SalamonRakshit2020}%
  \BibitemOpen
  \bibfield  {author} {\bibinfo {author} {\bibfnamefont {T.}~\bibnamefont
  {Salamon}}, \bibinfo {author} {\bibfnamefont {A.}~\bibnamefont {Celi}},
  \bibinfo {author} {\bibfnamefont {R.~W.}\ \bibnamefont {Chhajlany}}, \bibinfo
  {author} {\bibfnamefont {I.}~\bibnamefont {Fr\'erot}}, \bibinfo {author}
  {\bibfnamefont {M.}~\bibnamefont {Lewenstein}}, \bibinfo {author}
  {\bibfnamefont {L.}~\bibnamefont {Tarruell}},\ and\ \bibinfo {author}
  {\bibfnamefont {D.}~\bibnamefont {Rakshit}},\ }\href
  {https://doi.org/https://doi.org/10.1103/PhysRevLett.125.030504} {\bibfield
  {journal} {\bibinfo  {journal} {Phys. Rev. Lett.}\ }\textbf {\bibinfo
  {volume} {125}},\ \bibinfo {pages} {030504} (\bibinfo {year}
  {2020})}\BibitemShut {NoStop}%
\bibitem [{\citenamefont {Chou}\ \emph {et~al.}(2020)\citenamefont {Chou},
  \citenamefont {Fu}, \citenamefont {Wilson}, \citenamefont {K\"onig},\ and\
  \citenamefont {Pixley}}]{ChouPixley2020}%
  \BibitemOpen
  \bibfield  {author} {\bibinfo {author} {\bibfnamefont {Y.-Z.}\ \bibnamefont
  {Chou}}, \bibinfo {author} {\bibfnamefont {Y.}~\bibnamefont {Fu}}, \bibinfo
  {author} {\bibfnamefont {J.~H.}\ \bibnamefont {Wilson}}, \bibinfo {author}
  {\bibfnamefont {E.~J.}\ \bibnamefont {K\"onig}},\ and\ \bibinfo {author}
  {\bibfnamefont {J.~H.}\ \bibnamefont {Pixley}},\ }\href
  {https://doi.org/https://doi.org/10.1103/PhysRevB.101.235121} {\bibfield
  {journal} {\bibinfo  {journal} {Phys. Rev. B}\ }\textbf {\bibinfo {volume}
  {101}},\ \bibinfo {pages} {235121} (\bibinfo {year} {2020})}\BibitemShut
  {NoStop}%
\bibitem [{\citenamefont {Mao}\ and\ \citenamefont
  {Senthil}(2021)}]{MaoSenthil2021}%
  \BibitemOpen
  \bibfield  {author} {\bibinfo {author} {\bibfnamefont {D.}~\bibnamefont
  {Mao}}\ and\ \bibinfo {author} {\bibfnamefont {T.}~\bibnamefont {Senthil}},\
  }\href {https://doi.org/https://doi.org/10.1103/PhysRevB.103.115110}
  {\bibfield  {journal} {\bibinfo  {journal} {Phys. Rev. B}\ }\textbf {\bibinfo
  {volume} {103}},\ \bibinfo {pages} {115110} (\bibinfo {year}
  {2021})}\BibitemShut {NoStop}%
\bibitem [{\citenamefont {Meng}\ \emph {et~al.}(2021)\citenamefont {Meng},
  \citenamefont {Wang}, \citenamefont {Han}, \citenamefont {Liu}, \citenamefont
  {Wen}, \citenamefont {Gao}, \citenamefont {Wang}, \citenamefont {Chin},\ and\
  \citenamefont {Zhang}}]{MengZhang2021}%
  \BibitemOpen
  \bibfield  {author} {\bibinfo {author} {\bibfnamefont {Z.}~\bibnamefont
  {Meng}}, \bibinfo {author} {\bibfnamefont {L.}~\bibnamefont {Wang}}, \bibinfo
  {author} {\bibfnamefont {W.}~\bibnamefont {Han}}, \bibinfo {author}
  {\bibfnamefont {F.}~\bibnamefont {Liu}}, \bibinfo {author} {\bibfnamefont
  {K.}~\bibnamefont {Wen}}, \bibinfo {author} {\bibfnamefont {C.}~\bibnamefont
  {Gao}}, \bibinfo {author} {\bibfnamefont {P.}~\bibnamefont {Wang}}, \bibinfo
  {author} {\bibfnamefont {C.}~\bibnamefont {Chin}},\ and\ \bibinfo {author}
  {\bibfnamefont {J.}~\bibnamefont {Zhang}},\ }\href
  {https://doi.org/https://doi.org/10.48550/ARXIV.2110.00149} {\bibinfo {title}
  {Atomic bose-einstein condensate in a twisted-bilayer optical lattice}}
  (\bibinfo {year} {2021}),\ \Eprint {https://arxiv.org/abs/2110.00149}
  {arXiv:2110.00149} \BibitemShut {NoStop}%
\bibitem [{\citenamefont {Lee}\ and\ \citenamefont
  {Pixley}(2022)}]{LeePixley2022}%
  \BibitemOpen
  \bibfield  {author} {\bibinfo {author} {\bibfnamefont {J.}~\bibnamefont
  {Lee}}\ and\ \bibinfo {author} {\bibfnamefont {J.~H.}\ \bibnamefont
  {Pixley}},\ }\href {https://doi.org/10.21468/SciPostPhys.13.2.033} {\bibfield
   {journal} {\bibinfo  {journal} {SciPost Phys.}\ }\textbf {\bibinfo {volume}
  {13}},\ \bibinfo {pages} {033} (\bibinfo {year} {2022})}\BibitemShut
  {NoStop}%
\bibitem [{\citenamefont {Landauer}(1976)}]{Landauer_1976}%
  \BibitemOpen
  \bibfield  {author} {\bibinfo {author} {\bibfnamefont {R.}~\bibnamefont
  {Landauer}},\ }\href@noop {} {\bibfield  {journal} {\bibinfo  {journal}
  {Collect. Phenom.}\ }\textbf {\bibinfo {volume} {2}},\ \bibinfo {pages} {167}
  (\bibinfo {year} {1976})}\BibitemShut {NoStop}%
\bibitem [{\citenamefont {Catalan}\ \emph {et~al.}(2015)\citenamefont
  {Catalan}, \citenamefont {Jim{\'e}nez},\ and\ \citenamefont
  {Gruverman}}]{Catalan_2015}%
  \BibitemOpen
  \bibfield  {author} {\bibinfo {author} {\bibfnamefont {G.}~\bibnamefont
  {Catalan}}, \bibinfo {author} {\bibfnamefont {D.}~\bibnamefont
  {Jim{\'e}nez}},\ and\ \bibinfo {author} {\bibfnamefont {A.}~\bibnamefont
  {Gruverman}},\ }\href {https://doi.org/https://doi.org/10.1038/nmat4195}
  {\bibfield  {journal} {\bibinfo  {journal} {Nature Materials}\ }\textbf
  {\bibinfo {volume} {14}},\ \bibinfo {pages} {137} (\bibinfo {year}
  {2015})}\BibitemShut {NoStop}%
\bibitem [{\citenamefont {Ng}\ \emph {et~al.}(2017)\citenamefont {Ng},
  \citenamefont {Hillenius},\ and\ \citenamefont {Gruverman}}]{Ng_2017}%
  \BibitemOpen
  \bibfield  {author} {\bibinfo {author} {\bibfnamefont {K.}~\bibnamefont
  {Ng}}, \bibinfo {author} {\bibfnamefont {S.~J.}\ \bibnamefont {Hillenius}},\
  and\ \bibinfo {author} {\bibfnamefont {A.}~\bibnamefont {Gruverman}},\ }\href
  {https://doi.org/https://doi.org/10.1016/j.ssc.2017.07.020} {\bibfield
  {journal} {\bibinfo  {journal} {Solid State Communications}\ }\textbf
  {\bibinfo {volume} {265}},\ \bibinfo {pages} {12 } (\bibinfo {year}
  {2017})}\BibitemShut {NoStop}%
\bibitem [{\citenamefont {Hoffmann}\ \emph {et~al.}(2018)\citenamefont
  {Hoffmann}, \citenamefont {Khan}, \citenamefont {Serrao}, \citenamefont {Lu},
  \citenamefont {Salahuddin}, \citenamefont {Pe\v{s}i\'{c}}, \citenamefont
  {Slesazeck}, \citenamefont {Schroeder},\ and\ \citenamefont
  {Mikolajick}}]{Hoffmann_2018}%
  \BibitemOpen
  \bibfield  {author} {\bibinfo {author} {\bibfnamefont {M.}~\bibnamefont
  {Hoffmann}}, \bibinfo {author} {\bibfnamefont {A.~I.}\ \bibnamefont {Khan}},
  \bibinfo {author} {\bibfnamefont {C.}~\bibnamefont {Serrao}}, \bibinfo
  {author} {\bibfnamefont {Z.}~\bibnamefont {Lu}}, \bibinfo {author}
  {\bibfnamefont {S.}~\bibnamefont {Salahuddin}}, \bibinfo {author}
  {\bibfnamefont {M.}~\bibnamefont {Pe\v{s}i\'{c}}}, \bibinfo {author}
  {\bibfnamefont {S.}~\bibnamefont {Slesazeck}}, \bibinfo {author}
  {\bibfnamefont {U.}~\bibnamefont {Schroeder}},\ and\ \bibinfo {author}
  {\bibfnamefont {T.}~\bibnamefont {Mikolajick}},\ }\href
  {https://doi.org/https://doi.org/10.1063/1.5030072} {\bibfield  {journal}
  {\bibinfo  {journal} {Journal of Applied Physics}\ }\textbf {\bibinfo
  {volume} {123}},\ \bibinfo {pages} {184101} (\bibinfo {year}
  {2018})}\BibitemShut {NoStop}%
\bibitem [{\citenamefont {Luk'yanchuk}\ \emph {et~al.}(2019)\citenamefont
  {Luk'yanchuk}, \citenamefont {Tikhonov}, \citenamefont {Sen{\'e}},
  \citenamefont {Razumnaya},\ and\ \citenamefont {Vinokur}}]{Lukyanchuk_2019}%
  \BibitemOpen
  \bibfield  {author} {\bibinfo {author} {\bibfnamefont {I.}~\bibnamefont
  {Luk'yanchuk}}, \bibinfo {author} {\bibfnamefont {Y.}~\bibnamefont
  {Tikhonov}}, \bibinfo {author} {\bibfnamefont {A.}~\bibnamefont {Sen{\'e}}},
  \bibinfo {author} {\bibfnamefont {A.}~\bibnamefont {Razumnaya}},\ and\
  \bibinfo {author} {\bibfnamefont {V.~M.}\ \bibnamefont {Vinokur}},\ }\href
  {https://doi.org/https://doi.org/10.1038/s42005-019-0121-0} {\bibfield
  {journal} {\bibinfo  {journal} {Communications Physics}\ }\textbf {\bibinfo
  {volume} {2}},\ \bibinfo {pages} {22} (\bibinfo {year} {2019})}\BibitemShut
  {NoStop}%
\bibitem [{\citenamefont {Hoffmann}\ \emph {et~al.}(2020)\citenamefont
  {Hoffmann}, \citenamefont {Slesazeck}, \citenamefont {Schroeder},\ and\
  \citenamefont {Mikolajick}}]{Hoffmann_2020}%
  \BibitemOpen
  \bibfield  {author} {\bibinfo {author} {\bibfnamefont {M.}~\bibnamefont
  {Hoffmann}}, \bibinfo {author} {\bibfnamefont {S.}~\bibnamefont {Slesazeck}},
  \bibinfo {author} {\bibfnamefont {U.}~\bibnamefont {Schroeder}},\ and\
  \bibinfo {author} {\bibfnamefont {T.}~\bibnamefont {Mikolajick}},\ }\href
  {https://doi.org/https://doi.org/10.1038/s41928-020-00474-9} {\bibfield
  {journal} {\bibinfo  {journal} {Nature Electronics}\ }\textbf {\bibinfo
  {volume} {3}},\ \bibinfo {pages} {504} (\bibinfo {year} {2020})}\BibitemShut
  {NoStop}%
\bibitem [{\citenamefont {Salahuddin}\ and\ \citenamefont
  {Datta}(2008)}]{Salahuddin_2008}%
  \BibitemOpen
  \bibfield  {author} {\bibinfo {author} {\bibfnamefont {S.}~\bibnamefont
  {Salahuddin}}\ and\ \bibinfo {author} {\bibfnamefont {S.}~\bibnamefont
  {Datta}},\ }\href {https://doi.org/https://doi.org/10.1021/nl071804g}
  {\bibfield  {journal} {\bibinfo  {journal} {Nano Letters}\ }\textbf {\bibinfo
  {volume} {8}},\ \bibinfo {pages} {405} (\bibinfo {year} {2008})}\BibitemShut
  {NoStop}%
\bibitem [{\citenamefont {Luryi}(1988)}]{Luryi_1988}%
  \BibitemOpen
  \bibfield  {author} {\bibinfo {author} {\bibfnamefont {S.}~\bibnamefont
  {Luryi}},\ }\href {https://doi.org/10.1063/1.99649} {\bibfield  {journal}
  {\bibinfo  {journal} {Applied Physics Letters}\ }\textbf {\bibinfo {volume}
  {52}},\ \bibinfo {pages} {501} (\bibinfo {year} {1988})}\BibitemShut
  {NoStop}%
\bibitem [{\citenamefont {Wang}\ \emph {et~al.}(2013)\citenamefont {Wang},
  \citenamefont {Wang}, \citenamefont {Chen}, \citenamefont {Zhu},
  \citenamefont {Zhu}, \citenamefont {Wu}, \citenamefont {Han}, \citenamefont
  {Zhang}, \citenamefont {Li}, \citenamefont {He}, \citenamefont {Xiong},
  \citenamefont {Law}, \citenamefont {Su},\ and\ \citenamefont
  {Wang}}]{Wang_2013}%
  \BibitemOpen
  \bibfield  {author} {\bibinfo {author} {\bibfnamefont {L.}~\bibnamefont
  {Wang}}, \bibinfo {author} {\bibfnamefont {Y.}~\bibnamefont {Wang}}, \bibinfo
  {author} {\bibfnamefont {X.}~\bibnamefont {Chen}}, \bibinfo {author}
  {\bibfnamefont {W.}~\bibnamefont {Zhu}}, \bibinfo {author} {\bibfnamefont
  {C.}~\bibnamefont {Zhu}}, \bibinfo {author} {\bibfnamefont {Z.}~\bibnamefont
  {Wu}}, \bibinfo {author} {\bibfnamefont {Y.}~\bibnamefont {Han}}, \bibinfo
  {author} {\bibfnamefont {M.}~\bibnamefont {Zhang}}, \bibinfo {author}
  {\bibfnamefont {W.}~\bibnamefont {Li}}, \bibinfo {author} {\bibfnamefont
  {Y.}~\bibnamefont {He}}, \bibinfo {author} {\bibfnamefont {W.}~\bibnamefont
  {Xiong}}, \bibinfo {author} {\bibfnamefont {K.~T.}\ \bibnamefont {Law}},
  \bibinfo {author} {\bibfnamefont {D.}~\bibnamefont {Su}},\ and\ \bibinfo
  {author} {\bibfnamefont {N.}~\bibnamefont {Wang}},\ }\href
  {https://doi.org/10.1038/srep02041} {\bibfield  {journal} {\bibinfo
  {journal} {Scientific Reports}\ }\textbf {\bibinfo {volume} {3}},\ \bibinfo
  {pages} {2041} (\bibinfo {year} {2013})}\BibitemShut {NoStop}%
\bibitem [{\citenamefont {Choi}\ \emph {et~al.}(2016)\citenamefont {Choi},
  \citenamefont {Lee}, \citenamefont {Park}, \citenamefont {Yu}, \citenamefont
  {Kim},\ and\ \citenamefont {Shin}}]{Choi_2016}%
  \BibitemOpen
  \bibfield  {author} {\bibinfo {author} {\bibfnamefont {H.}~\bibnamefont
  {Choi}}, \bibinfo {author} {\bibfnamefont {H.}~\bibnamefont {Lee}}, \bibinfo
  {author} {\bibfnamefont {J.}~\bibnamefont {Park}}, \bibinfo {author}
  {\bibfnamefont {H.-Y.}\ \bibnamefont {Yu}}, \bibinfo {author} {\bibfnamefont
  {T.~G.}\ \bibnamefont {Kim}},\ and\ \bibinfo {author} {\bibfnamefont
  {C.}~\bibnamefont {Shin}},\ }\href {https://doi.org/10.1063/1.4968183}
  {\bibfield  {journal} {\bibinfo  {journal} {Applied Physics Letters}\
  }\textbf {\bibinfo {volume} {109}},\ \bibinfo {pages} {203505} (\bibinfo
  {year} {2016})}\BibitemShut {NoStop}%
\bibitem [{\citenamefont {Giordano}(1988)}]{Giordano1988}%
  \BibitemOpen
  \bibfield  {author} {\bibinfo {author} {\bibfnamefont {N.}~\bibnamefont
  {Giordano}},\ }\href
  {https://doi.org/https://doi.org/10.1103/PhysRevLett.61.2137} {\bibfield
  {journal} {\bibinfo  {journal} {Phys. Rev. Lett.}\ }\textbf {\bibinfo
  {volume} {61}},\ \bibinfo {pages} {2137} (\bibinfo {year}
  {1988})}\BibitemShut {NoStop}%
\bibitem [{\citenamefont {Bezryadin}\ \emph {et~al.}(2000)\citenamefont
  {Bezryadin}, \citenamefont {Lau},\ and\ \citenamefont
  {Tinkham}}]{Bezryadin2000}%
  \BibitemOpen
  \bibfield  {author} {\bibinfo {author} {\bibfnamefont {A.}~\bibnamefont
  {Bezryadin}}, \bibinfo {author} {\bibfnamefont {C.}~\bibnamefont {Lau}},\
  and\ \bibinfo {author} {\bibfnamefont {M.}~\bibnamefont {Tinkham}},\ }\href
  {https://doi.org/https://doi.org/10.1038/35010060} {\bibfield  {journal}
  {\bibinfo  {journal} {Nature}\ }\textbf {\bibinfo {volume} {404}},\ \bibinfo
  {pages} {971} (\bibinfo {year} {2000})}\BibitemShut {NoStop}%
\bibitem [{\citenamefont {Lau}\ \emph {et~al.}(2001)\citenamefont {Lau},
  \citenamefont {Markovic}, \citenamefont {Bockrath}, \citenamefont
  {Bezryadin},\ and\ \citenamefont {Tinkham}}]{Lau_2001}%
  \BibitemOpen
  \bibfield  {author} {\bibinfo {author} {\bibfnamefont {C.~N.}\ \bibnamefont
  {Lau}}, \bibinfo {author} {\bibfnamefont {N.}~\bibnamefont {Markovic}},
  \bibinfo {author} {\bibfnamefont {M.}~\bibnamefont {Bockrath}}, \bibinfo
  {author} {\bibfnamefont {A.}~\bibnamefont {Bezryadin}},\ and\ \bibinfo
  {author} {\bibfnamefont {M.}~\bibnamefont {Tinkham}},\ }\href
  {https://doi.org/https://doi.org/10.1103/PhysRevLett.87.217003} {\bibfield
  {journal} {\bibinfo  {journal} {Phys. Rev. Lett.}\ }\textbf {\bibinfo
  {volume} {87}},\ \bibinfo {pages} {217003} (\bibinfo {year}
  {2001})}\BibitemShut {NoStop}%
\bibitem [{\citenamefont {B\"uchler}\ \emph {et~al.}(2004)\citenamefont
  {B\"uchler}, \citenamefont {Geshkenbein},\ and\ \citenamefont
  {Blatter}}]{Buchler2004}%
  \BibitemOpen
  \bibfield  {author} {\bibinfo {author} {\bibfnamefont {H.~P.}\ \bibnamefont
  {B\"uchler}}, \bibinfo {author} {\bibfnamefont {V.~B.}\ \bibnamefont
  {Geshkenbein}},\ and\ \bibinfo {author} {\bibfnamefont {G.}~\bibnamefont
  {Blatter}},\ }\href
  {https://doi.org/https://doi.org/10.1103/PhysRevLett.92.067007} {\bibfield
  {journal} {\bibinfo  {journal} {Phys. Rev. Lett.}\ }\textbf {\bibinfo
  {volume} {92}},\ \bibinfo {pages} {067007} (\bibinfo {year}
  {2004})}\BibitemShut {NoStop}%
\bibitem [{\citenamefont {Mooij}\ and\ \citenamefont
  {Harmans}(2005)}]{Mooij_2005}%
  \BibitemOpen
  \bibfield  {author} {\bibinfo {author} {\bibfnamefont {J.~E.}\ \bibnamefont
  {Mooij}}\ and\ \bibinfo {author} {\bibfnamefont {C.~J. P.~M.}\ \bibnamefont
  {Harmans}},\ }\href
  {https://doi.org/https://doi.org/10.1088/1367-2630/7/1/219} {\bibfield
  {journal} {\bibinfo  {journal} {New Journal of Physics}\ }\textbf {\bibinfo
  {volume} {7}},\ \bibinfo {pages} {219} (\bibinfo {year} {2005})}\BibitemShut
  {NoStop}%
\bibitem [{\citenamefont {Mooij}\ and\ \citenamefont
  {Nazarov}(2006)}]{Mooij_2006}%
  \BibitemOpen
  \bibfield  {author} {\bibinfo {author} {\bibfnamefont {J.~E.}\ \bibnamefont
  {Mooij}}\ and\ \bibinfo {author} {\bibfnamefont {Y.~V.}\ \bibnamefont
  {Nazarov}},\ }\href {https://doi.org/https://doi.org/10.1038/nphys234}
  {\bibfield  {journal} {\bibinfo  {journal} {Nature Physics}\ }\textbf
  {\bibinfo {volume} {2}},\ \bibinfo {pages} {169} (\bibinfo {year}
  {2006})}\BibitemShut {NoStop}%
\bibitem [{\citenamefont {Arutyunov}\ \emph {et~al.}(2008)\citenamefont
  {Arutyunov}, \citenamefont {Golubev},\ and\ \citenamefont
  {Zaikin}}]{Arutyunov2008}%
  \BibitemOpen
  \bibfield  {author} {\bibinfo {author} {\bibfnamefont {K.}~\bibnamefont
  {Arutyunov}}, \bibinfo {author} {\bibfnamefont {D.}~\bibnamefont {Golubev}},\
  and\ \bibinfo {author} {\bibfnamefont {A.}~\bibnamefont {Zaikin}},\ }\href
  {https://doi.org/https://doi.org/10.1016/j.physrep.2008.04.009} {\bibfield
  {journal} {\bibinfo  {journal} {Physics Reports}\ }\textbf {\bibinfo {volume}
  {464}},\ \bibinfo {pages} {1} (\bibinfo {year} {2008})}\BibitemShut {NoStop}%
\bibitem [{\citenamefont {Astafiev}\ \emph {et~al.}(2012)\citenamefont
  {Astafiev}, \citenamefont {Ioffe}, \citenamefont {Kafanov}, \citenamefont
  {Pashkin}, \citenamefont {Arutyunov}, \citenamefont {Shahar}, \citenamefont
  {Cohen},\ and\ \citenamefont {Tsai}}]{Astafiev_2012}%
  \BibitemOpen
  \bibfield  {author} {\bibinfo {author} {\bibfnamefont {O.~V.}\ \bibnamefont
  {Astafiev}}, \bibinfo {author} {\bibfnamefont {L.~B.}\ \bibnamefont {Ioffe}},
  \bibinfo {author} {\bibfnamefont {S.}~\bibnamefont {Kafanov}}, \bibinfo
  {author} {\bibfnamefont {Y.~A.}\ \bibnamefont {Pashkin}}, \bibinfo {author}
  {\bibfnamefont {K.~Y.}\ \bibnamefont {Arutyunov}}, \bibinfo {author}
  {\bibfnamefont {D.}~\bibnamefont {Shahar}}, \bibinfo {author} {\bibfnamefont
  {O.}~\bibnamefont {Cohen}},\ and\ \bibinfo {author} {\bibfnamefont {J.~S.}\
  \bibnamefont {Tsai}},\ }\href
  {https://doi.org/https://doi.org/10.1038/nature10930} {\bibfield  {journal}
  {\bibinfo  {journal} {Nature}\ }\textbf {\bibinfo {volume} {484}},\ \bibinfo
  {pages} {355} (\bibinfo {year} {2012})}\BibitemShut {NoStop}%
\bibitem [{\citenamefont {Ulrich}\ and\ \citenamefont
  {Hassler}(2016)}]{Ulrich_2016}%
  \BibitemOpen
  \bibfield  {author} {\bibinfo {author} {\bibfnamefont {J.}~\bibnamefont
  {Ulrich}}\ and\ \bibinfo {author} {\bibfnamefont {F.}~\bibnamefont
  {Hassler}},\ }\href
  {https://doi.org/https://doi.org/10.1103/PhysRevB.94.094505} {\bibfield
  {journal} {\bibinfo  {journal} {Phys. Rev. B}\ }\textbf {\bibinfo {volume}
  {94}},\ \bibinfo {pages} {094505} (\bibinfo {year} {2016})}\BibitemShut
  {NoStop}%
\bibitem [{\citenamefont {de~Graaf}\ \emph {et~al.}(2018)\citenamefont
  {de~Graaf}, \citenamefont {Skacel}, \citenamefont {H{\"o}nigl-Decrinis},
  \citenamefont {Shaikhaidarov}, \citenamefont {Rotzinger}, \citenamefont
  {Linzen}, \citenamefont {Ziegler}, \citenamefont {H{\"u}bner}, \citenamefont
  {Meyer}, \citenamefont {Antonov}, \citenamefont {Il'ichev}, \citenamefont
  {Ustinov}, \citenamefont {Tzalenchuk},\ and\ \citenamefont
  {Astafiev}}]{deGraaf_2018}%
  \BibitemOpen
  \bibfield  {author} {\bibinfo {author} {\bibfnamefont {S.~E.}\ \bibnamefont
  {de~Graaf}}, \bibinfo {author} {\bibfnamefont {S.~T.}\ \bibnamefont
  {Skacel}}, \bibinfo {author} {\bibfnamefont {T.}~\bibnamefont
  {H{\"o}nigl-Decrinis}}, \bibinfo {author} {\bibfnamefont {R.}~\bibnamefont
  {Shaikhaidarov}}, \bibinfo {author} {\bibfnamefont {H.}~\bibnamefont
  {Rotzinger}}, \bibinfo {author} {\bibfnamefont {S.}~\bibnamefont {Linzen}},
  \bibinfo {author} {\bibfnamefont {M.}~\bibnamefont {Ziegler}}, \bibinfo
  {author} {\bibfnamefont {U.}~\bibnamefont {H{\"u}bner}}, \bibinfo {author}
  {\bibfnamefont {H.~G.}\ \bibnamefont {Meyer}}, \bibinfo {author}
  {\bibfnamefont {V.}~\bibnamefont {Antonov}}, \bibinfo {author} {\bibfnamefont
  {E.}~\bibnamefont {Il'ichev}}, \bibinfo {author} {\bibfnamefont {A.~V.}\
  \bibnamefont {Ustinov}}, \bibinfo {author} {\bibfnamefont {A.~Y.}\
  \bibnamefont {Tzalenchuk}},\ and\ \bibinfo {author} {\bibfnamefont {O.~V.}\
  \bibnamefont {Astafiev}},\ }\href
  {https://doi.org/https://doi.org/10.1038/s41567-018-0097-9} {\bibfield
  {journal} {\bibinfo  {journal} {Nature Physics}\ }\textbf {\bibinfo {volume}
  {14}},\ \bibinfo {pages} {590} (\bibinfo {year} {2018})}\BibitemShut
  {NoStop}%
\bibitem [{\citenamefont {Li}\ \emph {et~al.}(2019)\citenamefont {Li},
  \citenamefont {Li}, \citenamefont {Lam},\ and\ \citenamefont
  {You}}]{Li_2019}%
  \BibitemOpen
  \bibfield  {author} {\bibinfo {author} {\bibfnamefont {Z.-Z.}\ \bibnamefont
  {Li}}, \bibinfo {author} {\bibfnamefont {T.-F.}\ \bibnamefont {Li}}, \bibinfo
  {author} {\bibfnamefont {C.-H.}\ \bibnamefont {Lam}},\ and\ \bibinfo {author}
  {\bibfnamefont {J.~Q.}\ \bibnamefont {You}},\ }\href
  {https://doi.org/https://doi.org/10.1103/PhysRevA.99.012309} {\bibfield
  {journal} {\bibinfo  {journal} {Phys. Rev. A}\ }\textbf {\bibinfo {volume}
  {99}},\ \bibinfo {pages} {012309} (\bibinfo {year} {2019})}\BibitemShut
  {NoStop}%
\bibitem [{\citenamefont {Shaikhaidarov}\ \emph {et~al.}(2022)\citenamefont
  {Shaikhaidarov}, \citenamefont {Kim}, \citenamefont {Dunstan}, \citenamefont
  {Antonov}, \citenamefont {Linzen}, \citenamefont {Ziegler}, \citenamefont
  {Golubev}, \citenamefont {Antonov}, \citenamefont {Il'ichev},\ and\
  \citenamefont {Astafiev}}]{Shaikhaidarov_2022}%
  \BibitemOpen
  \bibfield  {author} {\bibinfo {author} {\bibfnamefont {R.~S.}\ \bibnamefont
  {Shaikhaidarov}}, \bibinfo {author} {\bibfnamefont {K.~H.}\ \bibnamefont
  {Kim}}, \bibinfo {author} {\bibfnamefont {J.~W.}\ \bibnamefont {Dunstan}},
  \bibinfo {author} {\bibfnamefont {I.~V.}\ \bibnamefont {Antonov}}, \bibinfo
  {author} {\bibfnamefont {S.}~\bibnamefont {Linzen}}, \bibinfo {author}
  {\bibfnamefont {M.}~\bibnamefont {Ziegler}}, \bibinfo {author} {\bibfnamefont
  {D.~S.}\ \bibnamefont {Golubev}}, \bibinfo {author} {\bibfnamefont {V.~N.}\
  \bibnamefont {Antonov}}, \bibinfo {author} {\bibfnamefont {E.~V.}\
  \bibnamefont {Il'ichev}},\ and\ \bibinfo {author} {\bibfnamefont {O.~V.}\
  \bibnamefont {Astafiev}},\ }\href
  {https://doi.org/10.1038/s41586-022-04947-z} {\bibfield  {journal} {\bibinfo
  {journal} {Nature}\ }\textbf {\bibinfo {volume} {608}},\ \bibinfo {pages}
  {45} (\bibinfo {year} {2022})}\BibitemShut {NoStop}%
\bibitem [{\citenamefont {Koliofoti}\ and\ \citenamefont
  {Riwar}()}]{koliofoti2022}%
  \BibitemOpen
  \bibfield  {author} {\bibinfo {author} {\bibfnamefont {C.}~\bibnamefont
  {Koliofoti}}\ and\ \bibinfo {author} {\bibfnamefont {R.-P.}\ \bibnamefont
  {Riwar}},\ }\href@noop {} {}\Eprint {https://arxiv.org/abs/2204.13633}
  {arXiv:2204.13633} \BibitemShut {NoStop}%
\bibitem [{\citenamefont {Riwar}\ and\ \citenamefont
  {DiVincenzo}(2022)}]{Riwar_2022}%
  \BibitemOpen
  \bibfield  {author} {\bibinfo {author} {\bibfnamefont {R.~P.}\ \bibnamefont
  {Riwar}}\ and\ \bibinfo {author} {\bibfnamefont {D.~P.}\ \bibnamefont
  {DiVincenzo}},\ }\href
  {https://doi.org/https://doi.org/10.1038/s41534-022-00539-x} {\bibfield
  {journal} {\bibinfo  {journal} {npj Quantum Information}\ }\textbf {\bibinfo
  {volume} {8}},\ \bibinfo {pages} {36} (\bibinfo {year} {2022})}\BibitemShut
  {NoStop}%
\bibitem [{\citenamefont {Larsen}\ \emph {et~al.}(2015)\citenamefont {Larsen},
  \citenamefont {Petersson}, \citenamefont {Kuemmeth}, \citenamefont
  {Jespersen}, \citenamefont {Krogstrup}, \citenamefont {Nyg\aa{}rd},\ and\
  \citenamefont {Marcus}}]{Larsen_2015}%
  \BibitemOpen
  \bibfield  {author} {\bibinfo {author} {\bibfnamefont {T.~W.}\ \bibnamefont
  {Larsen}}, \bibinfo {author} {\bibfnamefont {K.~D.}\ \bibnamefont
  {Petersson}}, \bibinfo {author} {\bibfnamefont {F.}~\bibnamefont {Kuemmeth}},
  \bibinfo {author} {\bibfnamefont {T.~S.}\ \bibnamefont {Jespersen}}, \bibinfo
  {author} {\bibfnamefont {P.}~\bibnamefont {Krogstrup}}, \bibinfo {author}
  {\bibfnamefont {J.}~\bibnamefont {Nyg\aa{}rd}},\ and\ \bibinfo {author}
  {\bibfnamefont {C.~M.}\ \bibnamefont {Marcus}},\ }\href
  {https://doi.org/10.1103/PhysRevLett.115.127001} {\bibfield  {journal}
  {\bibinfo  {journal} {Phys. Rev. Lett.}\ }\textbf {\bibinfo {volume} {115}},\
  \bibinfo {pages} {127001} (\bibinfo {year} {2015})}\BibitemShut {NoStop}%
\bibitem [{\citenamefont {de~Lange}\ \emph {et~al.}(2015)\citenamefont
  {de~Lange}, \citenamefont {van Heck}, \citenamefont {Bruno}, \citenamefont
  {van Woerkom}, \citenamefont {Geresdi}, \citenamefont {Plissard},
  \citenamefont {Bakkers}, \citenamefont {Akhmerov},\ and\ \citenamefont
  {DiCarlo}}]{deLange_2015}%
  \BibitemOpen
  \bibfield  {author} {\bibinfo {author} {\bibfnamefont {G.}~\bibnamefont
  {de~Lange}}, \bibinfo {author} {\bibfnamefont {B.}~\bibnamefont {van Heck}},
  \bibinfo {author} {\bibfnamefont {A.}~\bibnamefont {Bruno}}, \bibinfo
  {author} {\bibfnamefont {D.~J.}\ \bibnamefont {van Woerkom}}, \bibinfo
  {author} {\bibfnamefont {A.}~\bibnamefont {Geresdi}}, \bibinfo {author}
  {\bibfnamefont {S.~R.}\ \bibnamefont {Plissard}}, \bibinfo {author}
  {\bibfnamefont {E.~P. A.~M.}\ \bibnamefont {Bakkers}}, \bibinfo {author}
  {\bibfnamefont {A.~R.}\ \bibnamefont {Akhmerov}},\ and\ \bibinfo {author}
  {\bibfnamefont {L.}~\bibnamefont {DiCarlo}},\ }\href
  {https://doi.org/10.1103/PhysRevLett.115.127002} {\bibfield  {journal}
  {\bibinfo  {journal} {Phys. Rev. Lett.}\ }\textbf {\bibinfo {volume} {115}},\
  \bibinfo {pages} {127002} (\bibinfo {year} {2015})}\BibitemShut {NoStop}%
\bibitem [{\citenamefont {Kjaergaard}\ \emph {et~al.}(2017)\citenamefont
  {Kjaergaard}, \citenamefont {Suominen}, \citenamefont {Nowak}, \citenamefont
  {Akhmerov}, \citenamefont {Shabani}, \citenamefont {Palmstr\o{}m},
  \citenamefont {Nichele},\ and\ \citenamefont {Marcus}}]{Kjaergaard_2017}%
  \BibitemOpen
  \bibfield  {author} {\bibinfo {author} {\bibfnamefont {M.}~\bibnamefont
  {Kjaergaard}}, \bibinfo {author} {\bibfnamefont {H.~J.}\ \bibnamefont
  {Suominen}}, \bibinfo {author} {\bibfnamefont {M.~P.}\ \bibnamefont {Nowak}},
  \bibinfo {author} {\bibfnamefont {A.~R.}\ \bibnamefont {Akhmerov}}, \bibinfo
  {author} {\bibfnamefont {J.}~\bibnamefont {Shabani}}, \bibinfo {author}
  {\bibfnamefont {C.~J.}\ \bibnamefont {Palmstr\o{}m}}, \bibinfo {author}
  {\bibfnamefont {F.}~\bibnamefont {Nichele}},\ and\ \bibinfo {author}
  {\bibfnamefont {C.~M.}\ \bibnamefont {Marcus}},\ }\href
  {https://doi.org/10.1103/PhysRevApplied.7.034029} {\bibfield  {journal}
  {\bibinfo  {journal} {Phys. Rev. Appl.}\ }\textbf {\bibinfo {volume} {7}},\
  \bibinfo {pages} {034029} (\bibinfo {year} {2017})}\BibitemShut {NoStop}%
\bibitem [{\citenamefont {Casparis}\ \emph {et~al.}(2018)\citenamefont
  {Casparis}, \citenamefont {Connolly}, \citenamefont {Kjaergaard},
  \citenamefont {Pearson}, \citenamefont {Kringh{\o}j}, \citenamefont {Larsen},
  \citenamefont {Kuemmeth}, \citenamefont {Wang}, \citenamefont {Thomas},
  \citenamefont {Gronin}, \citenamefont {Gardner}, \citenamefont {Manfra},
  \citenamefont {Marcus},\ and\ \citenamefont {Petersson}}]{Casparis_2018}%
  \BibitemOpen
  \bibfield  {author} {\bibinfo {author} {\bibfnamefont {L.}~\bibnamefont
  {Casparis}}, \bibinfo {author} {\bibfnamefont {M.~R.}\ \bibnamefont
  {Connolly}}, \bibinfo {author} {\bibfnamefont {M.}~\bibnamefont
  {Kjaergaard}}, \bibinfo {author} {\bibfnamefont {N.~J.}\ \bibnamefont
  {Pearson}}, \bibinfo {author} {\bibfnamefont {A.}~\bibnamefont
  {Kringh{\o}j}}, \bibinfo {author} {\bibfnamefont {T.~W.}\ \bibnamefont
  {Larsen}}, \bibinfo {author} {\bibfnamefont {F.}~\bibnamefont {Kuemmeth}},
  \bibinfo {author} {\bibfnamefont {T.}~\bibnamefont {Wang}}, \bibinfo {author}
  {\bibfnamefont {C.}~\bibnamefont {Thomas}}, \bibinfo {author} {\bibfnamefont
  {S.}~\bibnamefont {Gronin}}, \bibinfo {author} {\bibfnamefont {G.~C.}\
  \bibnamefont {Gardner}}, \bibinfo {author} {\bibfnamefont {M.~J.}\
  \bibnamefont {Manfra}}, \bibinfo {author} {\bibfnamefont {C.~M.}\
  \bibnamefont {Marcus}},\ and\ \bibinfo {author} {\bibfnamefont {K.~D.}\
  \bibnamefont {Petersson}},\ }\href
  {https://doi.org/10.1038/s41565-018-0207-y} {\bibfield  {journal} {\bibinfo
  {journal} {Nature Nanotechnology}\ }\textbf {\bibinfo {volume} {13}},\
  \bibinfo {pages} {915} (\bibinfo {year} {2018})}\BibitemShut {NoStop}%
\bibitem [{\citenamefont {Lee}\ \emph {et~al.}(2019)\citenamefont {Lee},
  \citenamefont {Shojaei}, \citenamefont {Pendharkar}, \citenamefont
  {McFadden}, \citenamefont {Kim}, \citenamefont {Suominen}, \citenamefont
  {Kjaergaard}, \citenamefont {Nichele}, \citenamefont {Zhang}, \citenamefont
  {Marcus},\ and\ \citenamefont {Palmstr{\o}m}}]{Lee_2019}%
  \BibitemOpen
  \bibfield  {author} {\bibinfo {author} {\bibfnamefont {J.~S.}\ \bibnamefont
  {Lee}}, \bibinfo {author} {\bibfnamefont {B.}~\bibnamefont {Shojaei}},
  \bibinfo {author} {\bibfnamefont {M.}~\bibnamefont {Pendharkar}}, \bibinfo
  {author} {\bibfnamefont {A.~P.}\ \bibnamefont {McFadden}}, \bibinfo {author}
  {\bibfnamefont {Y.}~\bibnamefont {Kim}}, \bibinfo {author} {\bibfnamefont
  {H.~J.}\ \bibnamefont {Suominen}}, \bibinfo {author} {\bibfnamefont
  {M.}~\bibnamefont {Kjaergaard}}, \bibinfo {author} {\bibfnamefont
  {F.}~\bibnamefont {Nichele}}, \bibinfo {author} {\bibfnamefont
  {H.}~\bibnamefont {Zhang}}, \bibinfo {author} {\bibfnamefont {C.~M.}\
  \bibnamefont {Marcus}},\ and\ \bibinfo {author} {\bibfnamefont {C.~J.}\
  \bibnamefont {Palmstr{\o}m}},\ }\href
  {https://doi.org/10.1021/acs.nanolett.9b00494} {\bibfield  {journal}
  {\bibinfo  {journal} {Nano Letters}\ }\textbf {\bibinfo {volume} {19}},\
  \bibinfo {pages} {3083} (\bibinfo {year} {2019})}\BibitemShut {NoStop}%
\bibitem [{\citenamefont {Graziano}\ \emph {et~al.}(2022)\citenamefont
  {Graziano}, \citenamefont {Gupta}, \citenamefont {Pendharkar}, \citenamefont
  {Dong}, \citenamefont {Dempsey}, \citenamefont {Palmstr{\o}m},\ and\
  \citenamefont {Pribiag}}]{Graziano_2022}%
  \BibitemOpen
  \bibfield  {author} {\bibinfo {author} {\bibfnamefont {G.~V.}\ \bibnamefont
  {Graziano}}, \bibinfo {author} {\bibfnamefont {M.}~\bibnamefont {Gupta}},
  \bibinfo {author} {\bibfnamefont {M.}~\bibnamefont {Pendharkar}}, \bibinfo
  {author} {\bibfnamefont {J.~T.}\ \bibnamefont {Dong}}, \bibinfo {author}
  {\bibfnamefont {C.~P.}\ \bibnamefont {Dempsey}}, \bibinfo {author}
  {\bibfnamefont {C.}~\bibnamefont {Palmstr{\o}m}},\ and\ \bibinfo {author}
  {\bibfnamefont {V.~S.}\ \bibnamefont {Pribiag}},\ }\href
  {https://doi.org/10.1038/s41467-022-33682-2} {\bibfield  {journal} {\bibinfo
  {journal} {Nature Communications}\ }\textbf {\bibinfo {volume} {13}},\
  \bibinfo {pages} {5933} (\bibinfo {year} {2022})}\BibitemShut {NoStop}%
\bibitem [{\citenamefont {Koch}\ \emph {et~al.}(2007)\citenamefont {Koch},
  \citenamefont {Yu}, \citenamefont {Gambetta}, \citenamefont {Houck},
  \citenamefont {Schuster}, \citenamefont {Majer}, \citenamefont {Blais},
  \citenamefont {Devoret}, \citenamefont {Girvin},\ and\ \citenamefont
  {Schoelkopf}}]{Koch_2007}%
  \BibitemOpen
  \bibfield  {author} {\bibinfo {author} {\bibfnamefont {J.}~\bibnamefont
  {Koch}}, \bibinfo {author} {\bibfnamefont {T.~M.}\ \bibnamefont {Yu}},
  \bibinfo {author} {\bibfnamefont {J.}~\bibnamefont {Gambetta}}, \bibinfo
  {author} {\bibfnamefont {A.~A.}\ \bibnamefont {Houck}}, \bibinfo {author}
  {\bibfnamefont {D.~I.}\ \bibnamefont {Schuster}}, \bibinfo {author}
  {\bibfnamefont {J.}~\bibnamefont {Majer}}, \bibinfo {author} {\bibfnamefont
  {A.}~\bibnamefont {Blais}}, \bibinfo {author} {\bibfnamefont {M.~H.}\
  \bibnamefont {Devoret}}, \bibinfo {author} {\bibfnamefont {S.~M.}\
  \bibnamefont {Girvin}},\ and\ \bibinfo {author} {\bibfnamefont {R.~J.}\
  \bibnamefont {Schoelkopf}},\ }\href
  {https://doi.org/https://doi.org/10.1103/PhysRevA.76.042319} {\bibfield
  {journal} {\bibinfo  {journal} {Phys. Rev. A}\ }\textbf {\bibinfo {volume}
  {76}},\ \bibinfo {pages} {042319} (\bibinfo {year} {2007})}\BibitemShut
  {NoStop}%
\bibitem [{\citenamefont {Likharev}\ and\ \citenamefont
  {Zorin}(1985)}]{Likharev_1985}%
  \BibitemOpen
  \bibfield  {author} {\bibinfo {author} {\bibfnamefont {K.~K.}\ \bibnamefont
  {Likharev}}\ and\ \bibinfo {author} {\bibfnamefont {A.~B.}\ \bibnamefont
  {Zorin}},\ }\href {https://doi.org/https://doi.org/10.1007/BF00683782}
  {\bibfield  {journal} {\bibinfo  {journal} {Journal of Low Temperature
  Physics}\ }\textbf {\bibinfo {volume} {59}},\ \bibinfo {pages} {347}
  (\bibinfo {year} {1985})}\BibitemShut {NoStop}%
\bibitem [{\citenamefont {Loss}\ and\ \citenamefont
  {Mullen}(1991)}]{Loss_1991}%
  \BibitemOpen
  \bibfield  {author} {\bibinfo {author} {\bibfnamefont {D.}~\bibnamefont
  {Loss}}\ and\ \bibinfo {author} {\bibfnamefont {K.}~\bibnamefont {Mullen}},\
  }\href {https://doi.org/10.1103/PhysRevA.43.2129} {\bibfield  {journal}
  {\bibinfo  {journal} {Phys. Rev. A}\ }\textbf {\bibinfo {volume} {43}},\
  \bibinfo {pages} {2129} (\bibinfo {year} {1991})}\BibitemShut {NoStop}%
\bibitem [{\citenamefont {Koch}\ \emph {et~al.}(2009)\citenamefont {Koch},
  \citenamefont {Manucharyan}, \citenamefont {Devoret},\ and\ \citenamefont
  {Glazman}}]{Koch_2009}%
  \BibitemOpen
  \bibfield  {author} {\bibinfo {author} {\bibfnamefont {J.}~\bibnamefont
  {Koch}}, \bibinfo {author} {\bibfnamefont {V.}~\bibnamefont {Manucharyan}},
  \bibinfo {author} {\bibfnamefont {M.~H.}\ \bibnamefont {Devoret}},\ and\
  \bibinfo {author} {\bibfnamefont {L.~I.}\ \bibnamefont {Glazman}},\ }\href
  {https://doi.org/https://doi.org/10.1103/PhysRevLett.103.217004} {\bibfield
  {journal} {\bibinfo  {journal} {Phys. Rev. Lett.}\ }\textbf {\bibinfo
  {volume} {103}},\ \bibinfo {pages} {217004} (\bibinfo {year}
  {2009})}\BibitemShut {NoStop}%
\bibitem [{\citenamefont {Mizel}\ and\ \citenamefont
  {Yanay}(2020)}]{Mizel_2020}%
  \BibitemOpen
  \bibfield  {author} {\bibinfo {author} {\bibfnamefont {A.}~\bibnamefont
  {Mizel}}\ and\ \bibinfo {author} {\bibfnamefont {Y.}~\bibnamefont {Yanay}},\
  }\href {https://doi.org/https://doi.org/10.1103/PhysRevB.102.014512}
  {\bibfield  {journal} {\bibinfo  {journal} {Phys. Rev. B}\ }\textbf {\bibinfo
  {volume} {102}},\ \bibinfo {pages} {014512} (\bibinfo {year}
  {2020})}\BibitemShut {NoStop}%
\bibitem [{\citenamefont {Thanh~Le}\ \emph {et~al.}(2020)\citenamefont
  {Thanh~Le}, \citenamefont {Cole},\ and\ \citenamefont {Stace}}]{Thanh_2020}%
  \BibitemOpen
  \bibfield  {author} {\bibinfo {author} {\bibfnamefont {D.}~\bibnamefont
  {Thanh~Le}}, \bibinfo {author} {\bibfnamefont {J.~H.}\ \bibnamefont {Cole}},\
  and\ \bibinfo {author} {\bibfnamefont {T.~M.}\ \bibnamefont {Stace}},\ }\href
  {https://doi.org/https://doi.org/10.1103/PhysRevResearch.2.013245} {\bibfield
   {journal} {\bibinfo  {journal} {Phys. Rev. Research}\ }\textbf {\bibinfo
  {volume} {2}},\ \bibinfo {pages} {013245} (\bibinfo {year}
  {2020})}\BibitemShut {NoStop}%
\bibitem [{\citenamefont {Murani}\ \emph {et~al.}(2020)\citenamefont {Murani},
  \citenamefont {Bourlet}, \citenamefont {le~Sueur}, \citenamefont {Portier},
  \citenamefont {Altimiras}, \citenamefont {Esteve}, \citenamefont {Grabert},
  \citenamefont {Stockburger}, \citenamefont {Ankerhold},\ and\ \citenamefont
  {Joyez}}]{Murani_2020}%
  \BibitemOpen
  \bibfield  {author} {\bibinfo {author} {\bibfnamefont {A.}~\bibnamefont
  {Murani}}, \bibinfo {author} {\bibfnamefont {N.}~\bibnamefont {Bourlet}},
  \bibinfo {author} {\bibfnamefont {H.}~\bibnamefont {le~Sueur}}, \bibinfo
  {author} {\bibfnamefont {F.}~\bibnamefont {Portier}}, \bibinfo {author}
  {\bibfnamefont {C.}~\bibnamefont {Altimiras}}, \bibinfo {author}
  {\bibfnamefont {D.}~\bibnamefont {Esteve}}, \bibinfo {author} {\bibfnamefont
  {H.}~\bibnamefont {Grabert}}, \bibinfo {author} {\bibfnamefont
  {J.}~\bibnamefont {Stockburger}}, \bibinfo {author} {\bibfnamefont
  {J.}~\bibnamefont {Ankerhold}},\ and\ \bibinfo {author} {\bibfnamefont
  {P.}~\bibnamefont {Joyez}},\ }\href
  {https://doi.org/https://doi.org/10.1103/PhysRevX.10.021003} {\bibfield
  {journal} {\bibinfo  {journal} {Phys. Rev. X}\ }\textbf {\bibinfo {volume}
  {10}},\ \bibinfo {pages} {021003} (\bibinfo {year} {2020})}\BibitemShut
  {NoStop}%
\bibitem [{\citenamefont {Hakonen}\ and\ \citenamefont
  {Sonin}(2021)}]{Hakonen_2021}%
  \BibitemOpen
  \bibfield  {author} {\bibinfo {author} {\bibfnamefont {P.~J.}\ \bibnamefont
  {Hakonen}}\ and\ \bibinfo {author} {\bibfnamefont {E.~B.}\ \bibnamefont
  {Sonin}},\ }\href
  {https://doi.org/https://doi.org/10.1103/PhysRevX.11.018001} {\bibfield
  {journal} {\bibinfo  {journal} {Phys. Rev. X}\ }\textbf {\bibinfo {volume}
  {11}},\ \bibinfo {pages} {018001} (\bibinfo {year} {2021})}\BibitemShut
  {NoStop}%
\bibitem [{\citenamefont {Murani}\ \emph {et~al.}(2021)\citenamefont {Murani},
  \citenamefont {Bourlet}, \citenamefont {le~Sueur}, \citenamefont {Portier},
  \citenamefont {Altimiras}, \citenamefont {Esteve}, \citenamefont {Grabert},
  \citenamefont {Stockburger}, \citenamefont {Ankerhold},\ and\ \citenamefont
  {Joyez}}]{Murani_2021}%
  \BibitemOpen
  \bibfield  {author} {\bibinfo {author} {\bibfnamefont {A.}~\bibnamefont
  {Murani}}, \bibinfo {author} {\bibfnamefont {N.}~\bibnamefont {Bourlet}},
  \bibinfo {author} {\bibfnamefont {H.}~\bibnamefont {le~Sueur}}, \bibinfo
  {author} {\bibfnamefont {F.}~\bibnamefont {Portier}}, \bibinfo {author}
  {\bibfnamefont {C.}~\bibnamefont {Altimiras}}, \bibinfo {author}
  {\bibfnamefont {D.}~\bibnamefont {Esteve}}, \bibinfo {author} {\bibfnamefont
  {H.}~\bibnamefont {Grabert}}, \bibinfo {author} {\bibfnamefont
  {J.}~\bibnamefont {Stockburger}}, \bibinfo {author} {\bibfnamefont
  {J.}~\bibnamefont {Ankerhold}},\ and\ \bibinfo {author} {\bibfnamefont
  {P.}~\bibnamefont {Joyez}},\ }\href
  {https://doi.org/https://doi.org/10.1103/PhysRevX.11.018002} {\bibfield
  {journal} {\bibinfo  {journal} {Phys. Rev. X}\ }\textbf {\bibinfo {volume}
  {11}},\ \bibinfo {pages} {018002} (\bibinfo {year} {2021})}\BibitemShut
  {NoStop}%
\bibitem [{\citenamefont {Riwar}(2021)}]{Roman2021}%
  \BibitemOpen
  \bibfield  {author} {\bibinfo {author} {\bibfnamefont {R.-P.}\ \bibnamefont
  {Riwar}},\ }\href
  {https://doi.org/https://doi.org/10.21468/scipostphys.10.4.093} {\bibfield
  {journal} {\bibinfo  {journal} {SciPost Physics}\ }\textbf {\bibinfo {volume}
  {10}},\ \bibinfo {pages} {093} (\bibinfo {year} {2021})}\BibitemShut
  {NoStop}%
\bibitem [{\citenamefont {Devoret}(2021)}]{Devoret_2021}%
  \BibitemOpen
  \bibfield  {author} {\bibinfo {author} {\bibfnamefont {M.~H.}\ \bibnamefont
  {Devoret}},\ }\href {https://doi.org/10.1007/s10948-020-05784-9} {\bibfield
  {journal} {\bibinfo  {journal} {Journal of Superconductivity and Novel
  Magnetism}\ }\textbf {\bibinfo {volume} {34}},\ \bibinfo {pages} {1633}
  (\bibinfo {year} {2021})}\BibitemShut {NoStop}%
\bibitem [{\citenamefont {Kaur}\ \emph {et~al.}(2021)\citenamefont {Kaur},
  \citenamefont {S\'epulcre}, \citenamefont {Roch}, \citenamefont {Snyman},
  \citenamefont {Florens},\ and\ \citenamefont {Bera}}]{Kaur_2021}%
  \BibitemOpen
  \bibfield  {author} {\bibinfo {author} {\bibfnamefont {K.}~\bibnamefont
  {Kaur}}, \bibinfo {author} {\bibfnamefont {T.}~\bibnamefont {S\'epulcre}},
  \bibinfo {author} {\bibfnamefont {N.}~\bibnamefont {Roch}}, \bibinfo {author}
  {\bibfnamefont {I.}~\bibnamefont {Snyman}}, \bibinfo {author} {\bibfnamefont
  {S.}~\bibnamefont {Florens}},\ and\ \bibinfo {author} {\bibfnamefont
  {S.}~\bibnamefont {Bera}},\ }\href
  {https://doi.org/https://doi.org/10.1103/PhysRevLett.127.237702} {\bibfield
  {journal} {\bibinfo  {journal} {Phys. Rev. Lett.}\ }\textbf {\bibinfo
  {volume} {127}},\ \bibinfo {pages} {237702} (\bibinfo {year}
  {2021})}\BibitemShut {NoStop}%
\bibitem [{\citenamefont {Kenawy}\ \emph {et~al.}(2022)\citenamefont {Kenawy},
  \citenamefont {Hassler},\ and\ \citenamefont {Riwar}}]{Kenawy_2022}%
  \BibitemOpen
  \bibfield  {author} {\bibinfo {author} {\bibfnamefont {A.}~\bibnamefont
  {Kenawy}}, \bibinfo {author} {\bibfnamefont {F.}~\bibnamefont {Hassler}},\
  and\ \bibinfo {author} {\bibfnamefont {R.-P.}\ \bibnamefont {Riwar}},\ }\href
  {https://doi.org/10.1103/PhysRevB.106.035430} {\bibfield  {journal} {\bibinfo
   {journal} {Phys. Rev. B}\ }\textbf {\bibinfo {volume} {106}},\ \bibinfo
  {pages} {035430} (\bibinfo {year} {2022})}\BibitemShut {NoStop}%
\bibitem [{\citenamefont {Masuki}\ \emph {et~al.}(2022)\citenamefont {Masuki},
  \citenamefont {Sudo}, \citenamefont {Oshikawa},\ and\ \citenamefont
  {Ashida}}]{Masuki_2022}%
  \BibitemOpen
  \bibfield  {author} {\bibinfo {author} {\bibfnamefont {K.}~\bibnamefont
  {Masuki}}, \bibinfo {author} {\bibfnamefont {H.}~\bibnamefont {Sudo}},
  \bibinfo {author} {\bibfnamefont {M.}~\bibnamefont {Oshikawa}},\ and\
  \bibinfo {author} {\bibfnamefont {Y.}~\bibnamefont {Ashida}},\ }\href
  {https://doi.org/10.1103/PhysRevLett.129.087001} {\bibfield  {journal}
  {\bibinfo  {journal} {Phys. Rev. Lett.}\ }\textbf {\bibinfo {volume} {129}},\
  \bibinfo {pages} {087001} (\bibinfo {year} {2022})}\BibitemShut {NoStop}%
\bibitem [{\citenamefont {Kuzmin}\ \emph {et~al.}()\citenamefont {Kuzmin},
  \citenamefont {Mehta}, \citenamefont {Grabon}, \citenamefont {Mencia},
  \citenamefont {Burshtein}, \citenamefont {Goldstein},\ and\ \citenamefont
  {Manucharyan}}]{kuzmin2023observation}%
  \BibitemOpen
  \bibfield  {author} {\bibinfo {author} {\bibfnamefont {R.}~\bibnamefont
  {Kuzmin}}, \bibinfo {author} {\bibfnamefont {N.}~\bibnamefont {Mehta}},
  \bibinfo {author} {\bibfnamefont {N.}~\bibnamefont {Grabon}}, \bibinfo
  {author} {\bibfnamefont {R.~A.}\ \bibnamefont {Mencia}}, \bibinfo {author}
  {\bibfnamefont {A.}~\bibnamefont {Burshtein}}, \bibinfo {author}
  {\bibfnamefont {M.}~\bibnamefont {Goldstein}},\ and\ \bibinfo {author}
  {\bibfnamefont {V.~E.}\ \bibnamefont {Manucharyan}},\ }\href@noop {}
  {}\Eprint {https://arxiv.org/abs/2304.05806} {arXiv:2304.05806} \BibitemShut
  {NoStop}%
\bibitem [{\citenamefont {Kashuba}\ \emph {et~al.}()\citenamefont {Kashuba},
  \citenamefont {Schmidt}, \citenamefont {Hassler}, \citenamefont {Haller},\
  and\ \citenamefont {Riwar}}]{kashuba2023counting}%
  \BibitemOpen
  \bibfield  {author} {\bibinfo {author} {\bibfnamefont {O.}~\bibnamefont
  {Kashuba}}, \bibinfo {author} {\bibfnamefont {T.~L.}\ \bibnamefont
  {Schmidt}}, \bibinfo {author} {\bibfnamefont {F.}~\bibnamefont {Hassler}},
  \bibinfo {author} {\bibfnamefont {A.}~\bibnamefont {Haller}},\ and\ \bibinfo
  {author} {\bibfnamefont {R.~P.}\ \bibnamefont {Riwar}},\ }\href@noop {}
  {}\Eprint {https://arxiv.org/abs/2305.15906} {arXiv:2305.15906} \BibitemShut
  {NoStop}%
\bibitem [{Note1()}]{Note1}%
  \BibitemOpen
  \bibinfo {note} {This is true if $N_2$ is rescaled such that a change of
  $N_2$ by $\pm 1$ corresponds to an induced offset charge of an individual
  Cooper pair on the transistor.}\BibitemShut {Stop}%
\bibitem [{\citenamefont {Viola}\ and\ \citenamefont
  {DiVincenzo}(2014)}]{Viola_2014}%
  \BibitemOpen
  \bibfield  {author} {\bibinfo {author} {\bibfnamefont {G.}~\bibnamefont
  {Viola}}\ and\ \bibinfo {author} {\bibfnamefont {D.~P.}\ \bibnamefont
  {DiVincenzo}},\ }\href {https://doi.org/10.1103/PhysRevX.4.021019} {\bibfield
   {journal} {\bibinfo  {journal} {Phys. Rev. X}\ }\textbf {\bibinfo {volume}
  {4}},\ \bibinfo {pages} {021019} (\bibinfo {year} {2014})}\BibitemShut
  {NoStop}%
\bibitem [{\citenamefont {Rymarz}\ \emph {et~al.}(2021)\citenamefont {Rymarz},
  \citenamefont {Bosco}, \citenamefont {Ciani},\ and\ \citenamefont
  {DiVincenzo}}]{Rymarz2021}%
  \BibitemOpen
  \bibfield  {author} {\bibinfo {author} {\bibfnamefont {M.}~\bibnamefont
  {Rymarz}}, \bibinfo {author} {\bibfnamefont {S.}~\bibnamefont {Bosco}},
  \bibinfo {author} {\bibfnamefont {A.}~\bibnamefont {Ciani}},\ and\ \bibinfo
  {author} {\bibfnamefont {D.~P.}\ \bibnamefont {DiVincenzo}},\ }\href
  {https://doi.org/https://doi.org/10.1103/PhysRevX.11.011032} {\bibfield
  {journal} {\bibinfo  {journal} {Phys. Rev. X}\ }\textbf {\bibinfo {volume}
  {11}},\ \bibinfo {pages} {011032} (\bibinfo {year} {2021})}\BibitemShut
  {NoStop}%
\end{thebibliography}%

\newpage\hbox{}\thispagestyle{empty}\newpage

\widetext
\begin{center}
\textbf{\large Supplemental Material: Discrete control of capacitance in quantum circuits}
\end{center}
\setcounter{equation}{0}
\setcounter{figure}{0}
\setcounter{table}{0}
\setcounter{page}{1}
\makeatletter
\renewcommand{\theequation}{S\arabic{equation}}
\renewcommand{\thefigure}{S\arabic{figure}}
\renewcommand{\bibnumfmt}[1]{[S#1]}
\renewcommand{\citenumfont}[1]{S#1}

\section{Computing energy and Berry curvature of charge island transistor}

In the main text, we depict two possible models for transistors in Fig.~\ref{fig:chern_cylinder}. In this section, we derive the energy and the Berry curvature for the ground state of the first model, shown in Fig.~\ref{fig:chern_cylinder}a,b. The Hamiltonian is given as
\begin{equation}\label{eq_HT_Herrig}
    H_\text{T}=e_C(\widehat{n}+ N_2)^2-e_{J0}\cos(\widehat{\varphi})-e_{J1}\cos(\phi_1-\widehat{\varphi})\ ,
\end{equation}
where $[\widehat{\varphi},\widehat{n}]=i$ is the pair of canonically conjugate Cooper pair number and superconducting phase on the central transistor island. Its charging energy is $e_C$. The gate charge $N_2$ is capacitively coupled to the central island, and it has been rescaled such that a change of $N_2$ by $1$ induces a charge displacement inside the transistor island by one Cooper pair charge. The two electric contacts are coupled via the Josephson junctions $e_{J0}$ and $e_{J1}$, to ground and node $1$, respectively.

To proceed, we reformulate the Hamiltonian as
\begin{equation}
    H_\text{T}=e_C(\widehat{n}+ N_2)^2-e_{J}(\phi_1)\cos\left[\widehat{\varphi}-\delta(\phi_1)\right]\ ,
\end{equation}
where $e_{J}=\sqrt{e_{J0}^2+e_{J1}^2+2e_{J0}e_{J1}\cos(\phi_1)}$ and $\tan(\delta)=e_{J1}\sin(\phi_1)/[e_{J0}+e_{J1}\cos(\phi_1)]$. In the following, we focus for simplicity on the case where $e_C<e_J$, such that we may approximate the above cosine simply as a parabolic potential,
\begin{equation}
    H_\text{T}\approx e_C(\widehat{n}+ N_2)^2+\frac{e_{J}(\phi_1)}{2}\left[\widehat{\varphi}-\delta(\phi_1)\right]^2-e_J(\phi_1)\ ,
\end{equation}
which is readily solved by means of ordinary boson ladder operators,
\begin{align}
    \widehat{n}+N_2&=\left[\frac{e_J(\phi_1)}{2e_C}\right]^{1/4}\frac{\widehat{a}+\widehat{a}^\dagger}{\sqrt{2}}\\
    \widehat{\varphi}-\delta(\phi_1)&=i\left[\frac{2e_C}{e_J(\phi_1)}\right]^{1/4}\frac{\widehat{a}-\widehat{a}^\dagger}{\sqrt{2}}\ .
\end{align}
The inverse relationship yields
\begin{equation}\label{eq_a}
    \widehat{a}=\left[\frac{2e_{C}}{e_{J}\left(\phi_{1}\right)}\right]^{\frac{1}{4}}\frac{\widehat{n}+N_{2}}{\sqrt{2}}-i\left[\frac{e_{J}\left(\phi_{1}\right)}{2e_{C}}\right]^{\frac{1}{4}}\frac{\widehat{\varphi}-\delta\left(\phi_{1}\right)}{\sqrt{2}}\ .
\end{equation}
For the Hamiltonian we get
\begin{equation}
    H_\text{T}\approx \sqrt{2e_C e_J(\phi_1)}\left(\widehat{a}^\dagger \widehat{a}+\frac{1}{2}\right)-e_J(\phi_1)\ .
\end{equation}
The ground state energy consequently depends only on $\phi_1$,
\begin{equation}
    \epsilon\approx -e_J(\phi_1)+\sqrt{e_Ce_J(\phi_1)/2}\ .
\end{equation}
The $N_2$-dependence is exponentially suppressed due to $e_C<e_J$. To compute the Berry curvature, it is convenient to start with
\begin{equation}
    \mathcal{B}=-2\text{Im}\left[\sum_{m>0}\frac{\left\langle 0\right|\partial_{\phi_{1}}H\left|m\right\rangle \left\langle m\right|\partial_{N_{2}}H\left|0\right\rangle }{\left(\epsilon_{0}-\epsilon_{m}\right)^{2}}\right]\ .
\end{equation}
Plugging in the explicit operator forms for $\partial_{\phi_1}H$ and $\partial_{N_2}H$, and assuming $e_C<e_J$, we obtain
\begin{align}
\mathcal{B}&\approx-4e_{C}e_{J}\partial_{\phi_{1}}\delta\,\text{Im}\left[\sum_{m>0}\frac{\left\langle 0\right|\left(\widehat{\varphi}-\delta\right)\left|m\right\rangle \left\langle m\right|\left(\widehat{n}+N_{2}\right)\left|0\right\rangle }{\left(\epsilon_{0}-\epsilon_{m}\right)^{2}}\right]\\\label{eq_Berry_charge_explicit}&\approx\partial_{\phi_{1}}\delta=\frac{e_{J1}^{2}+e_{J0}e_{J1}\cos\left(\phi_{1}\right)}{e_{J1}^{2}+e_{J0}^{2}+2e_{J0}e_{J1}\cos\left(\phi_{1}\right)}\ ,
\end{align}
which is likewise independent of $N_2$.
For the Chern number we get $\mathcal{C}=0,1$ depending on whether $e_{J0}>e_{J1}$ or vice versa, as pointed out in the main text.

A final relevant result appearing in the main text is the leading order contribution to a nonflat $\epsilon$ and $\mathcal{B}$ in the cases $e_{J0}\gg e_{J1}$ and $e_{J0}\ll e_{J1}$. For $e_{J0}\gg e_{J1}$, we get up to first order in $e_{J1}$
\begin{align}\label{eq_eps_approx_0}
    \epsilon &\approx  -e_{J0}\left(1-\sqrt{\frac{e_C}{2e_{J0}}}\right)-e_{J1}\left(1-\sqrt{\frac{e_C}{8e_{J0}}}\right)\cos(\phi_1)\\\label{eq_B_approx_0}
    \mathcal{B} &\approx \frac{e_{J1}}{e_{J0}}\cos(\phi_1)\ . 
\end{align}
For $e_{J0}\ll e_{J1}$ on the other hand, we get up to first order in $e_{J0}$
\begin{align}\label{eq_eps_approx_1}
    \epsilon &\approx -e_{J1}\left(1-\sqrt{\frac{e_C}{2e_{J1}}}\right)-e_{J0}\left(1-\sqrt{\frac{e_C}{8e_{J1}}}\right)\cos(\phi_1)\\\label{eq_B_approx_1}
    \mathcal{B} &\approx 1-\frac{e_{J0}}{e_{J1}}\cos(\phi_1)\ . 
\end{align}
Consequently, as stated in the main text, the leading order contribution to a nonflat energy and Berry curvature, the term $\sim\cos(\phi_1)$, is always scaling linearly with the energy of the weaker junction.


\section{Computing energy and Berry curvature of quantum dot transistor}

In the main text, we depict a second transistor model with a quantum dot (Fig.~\ref{fig:chern_cylinder}c). In this section, we derive the energy and the Berry curvature for its many-body ground state. To describe this transistor realization, we deploy the Hamiltonian
\begin{equation}\label{eq_HT_QD}
\begin{split}
    H_\text{T}^\prime=\sum_{q\sigma}\left[\frac{\pi^{2}q^{2}}{2m_{e}L^{2}}-\mu+E_{\text{Th}}N_{2}\right]\widehat{c}_{q\sigma}^{\dagger}\widehat{c}_{q\sigma}\\+\sum_{q}\left(\frac{\Gamma_{0}+e^{i\phi_{1}}\Gamma_{1}}{2}\widehat{c}_{q\downarrow}\widehat{c}_{q\uparrow}+\text{h.c.}\right) \ ,
\end{split}
\end{equation}
where $\widehat{c}_{q\sigma}$ annihilates an electron with mass $m_e$ on the dot in level $q$ with spin $\sigma$, and $\Gamma_{0,1}$ describe the proximity effect between the quantum dot and the superconducting leads. The level spacing corresponds to the one of a 1D particle in a box of length $L$. We chose the 1D case simply for convenience, as it allows for a well-defined (convergent) computation of ground state energy without the need of additional cutoffs. The 2D and 3D cases work in close analogy and yield qualitatively the same results, but require the additional introduction of a Debye frequency cutoff for the pairing $\sim\Gamma_{0,1}$ for convergence, as is standard for BCS theory. $N_2$ has yet again been rescaled to correspond to the change of a Cooper pair charge, which is why the Thouless energy $E_\text{Th}=\pi v_F/L$ ($v_F=\sqrt{2\mu/m_{e}}$ is the Fermi velocity) appears as a prefactor (this detail will be explained in more detail below). 

The Hamiltonian can be cast into the Bogoliubov-de-Gennes
form
\begin{equation}
H_\text{T}^\prime=\sum_{q}\left(\widehat{c}_{q}^{\dagger}\mathcal{H}_{q}\widehat{c}_{q}+E_{q}\right)
\end{equation}
with
\begin{equation}
\widehat{c}_{q}^{\dagger}=\left(\widehat{c}_{q\uparrow}^{\dagger},\widehat{c}_{q\downarrow}\right)
\end{equation}
and
\begin{equation}\label{eq_HT_QD_BdG}
\mathcal{H}_{q}=\left(\begin{array}{cc}
\frac{\pi^2 q^2}{2m_{e}L^2L^2}-\mu+eV_{2} & \frac{\Gamma_0+e^{-i\phi_{1}}\Gamma_1}{2}\\
\frac{\Gamma_0+e^{i\phi_{1}}\Gamma_1}{2} & -\frac{\pi^2 q^2}{2m_{e}L^2L^2}+\mu-eV_{2}
\end{array}\right),
\end{equation}
as well as
\begin{equation}
E_{q}=\frac{\pi^{2}q^2}{2m_{e}L^2L^2}-\mu+eV_{2}
\end{equation}
which comes from the commutation of $\widehat{c}_{q\downarrow}^{\dagger}\widehat{c}_{q\downarrow}\rightarrow1-\widehat{c}_{q\downarrow}\widehat{c}_{q\downarrow}^{\dagger}$. Note that in Eq.~\eqref{eq_HT_QD_BdG} we defined the offset due to node $2$ in terms of the applied voltage, $eV_2$, in contrast to Eq.~\eqref{eq_HT_QD} where it is defined in terms of the charge on node $2$, $eV_2=E_\text{Th}N_2$. In this section, we first start the calculation with $eV_2$ and subsequently identify $E_\text{Th}$ as the relevant energy scale prefactor in front of $N_2$ based on how we defined $N_2$ in the main text: as the offset charge rescaled such that a change of $N_2$ by $\pm 1$ corresponds to a Cooper pair added to or removed from the quantum dot.

The single particle Hamiltonian $\mathcal{H}_{q}$ has eigenvalues
\begin{equation}
\epsilon_{q\pm}=\pm\sqrt{\left[\frac{\pi^{2}q^{2}}{2m_{e}L^2L^2}-\mu+eV_{2}\right]^{2}+\left|\frac{\Gamma_0+\Gamma_1e^{i\phi_{1}}}{2}\right|^{2}}
\end{equation}
with eigenvectors
\begin{equation}
\left(\begin{array}{c}
u_{q\pm}\\
v_{q\pm}
\end{array}\right)=\frac{1}{\sqrt{2}}\left(\begin{array}{c}
\sqrt{1\pm\frac{\frac{\pi^2 q^2}{2m_{e}L^2}-\mu+eV_{2}}{\sqrt{\left[\frac{\pi^2 q^2}{2m_{e}L^2}-\mu+eV_{2}\right]^{2}+\left|\frac{\Gamma_0+\Gamma_1e^{i\phi_{1}}}{2}\right|^{2}}}}\\
\pm\frac{\frac{\Gamma_0+\Gamma_1e^{i\phi_{1}}}{2}}{\left|\frac{\Gamma_0+\Gamma_1e^{i\phi_{1}}}{2}\right|}\sqrt{1\mp\frac{\frac{\pi^2 q^2}{2m_{e}L^2}-\mu+eV_{2}}{\sqrt{\left[\frac{\pi^2 q^2}{2m_{e}L^2}-\mu+eV_{2}\right]^{2}+\left|\frac{\Gamma_0+\Gamma_1e^{i\phi_{1}}}{2}\right|^{2}}}}
\end{array}\right).
\end{equation}
The many-body ground state energy is given as
\begin{equation}
\epsilon=\sum_{q}\left(\epsilon_{q-}+E_{q}\right).
\end{equation}
Note that while the quantum dot energy spectrum is discrete, we perform the sum for simplicity in the continuum
limit, $q\pi/L\rightarrow k$. This is a good approximation for $E_\text{Th}<\Gamma_{0,1}$. We get
\begin{align*}
\epsilon & =\frac{L}{\pi}\int_{0}^{\infty}dk\left[-\sqrt{\left[\frac{k^{2}}{2m_{e}}-\mu+eV_{2}\right]^{2}+\left|\frac{\Gamma_0+\Gamma_1e^{i\phi_{1}}}{2}\right|^{2}}+\frac{k^{2}}{2m_{e}}-\mu+eV_{2}\right]\\
 & =\frac{L}{2\pi}\sqrt{2m_{e}}\int_{0}^{\infty}d\omega\frac{1}{\sqrt{\omega}}\left[-\sqrt{\left[\omega-\underbrace{\left(\mu-eV_{2}\right)}_{\widetilde{\mu}}\right]^{2}+\underbrace{\left|\frac{\Gamma_0+\Gamma_1e^{i\phi_{1}}}{2}\right|^{2}}_{\left|\delta\right|^{2}}}+\omega-\underbrace{\left(\mu-eV_{2}\right)}_{\widetilde{\mu}}\right]
\end{align*}
Since $\widetilde{\mu}\gg\delta$, this integral is best solved in
three parts, by introducing a cutoff frequency $\omega_{\text{co}}$,
such that $\widetilde{\mu}\gg\omega_{\text{co}}\gg\delta$. We obtain
\begin{align*}
\int_{0}^{\infty}d\omega\frac{1}{\sqrt{\omega}}\left[\omega-\widetilde{\mu}-\sqrt{\left(\omega-\widetilde{\mu}\right)^{2}+\left|\delta\right|^{2}}\right] & \approx\int_{0}^{\widetilde{\mu}-\omega_{\text{co}}}d\omega\frac{1}{\sqrt{\omega}}\left[2\left(\omega-\widetilde{\mu}\right)-\frac{\left|\delta\right|^{2}}{2\left(\widetilde{\mu}-\omega\right)}\right]\\
 & +\int_{\widetilde{\mu}-\omega_{\text{co}}}^{\widetilde{\mu}+\omega_{\text{co}}}d\omega\frac{1}{\sqrt{\widetilde{\mu}}}\left[\omega-\widetilde{\mu}-\sqrt{\left(\omega-\widetilde{\mu}\right)^{2}+\left|\delta\right|^{2}}\right]\\
 & +\int_{\widetilde{\mu}+\omega_{\text{co}}}^{\infty}d\omega\frac{1}{\sqrt{\omega}}\left[-\frac{\left|\delta\right|^{2}}{2\left(\omega-\widetilde{\mu}\right)}\right].
\end{align*}
We thus end up at
\begin{align}
\int_{0}^{\infty}d\omega\frac{1}{\sqrt{\omega}}\left[\omega-\widetilde{\mu}-\sqrt{\left(\omega-\widetilde{\mu}\right)^{2}+\delta^{2}}\right] & \approx\frac{4}{3}\sqrt{\widetilde{\mu}-\omega_{\text{co}}}\left(-\omega_{\text{co}}-2\widetilde{\mu}\right)-\frac{\left|\delta\right|^{2}}{2\sqrt{\widetilde{\mu}}}\ln\left(\frac{1+\sqrt{\frac{\widetilde{\mu}-\omega_{\text{co}}}{\widetilde{\mu}}}}{1-\sqrt{\frac{\widetilde{\mu}-\omega_{\text{co}}}{\widetilde{\mu}}}}\right)\\
 & +\frac{-2\omega_{\text{co}}\sqrt{\left|\delta\right|^{2}+\omega_{\text{co}}^{2}}-\delta^{2}\ln\left(\frac{\sqrt{\left|\delta\right|^{2}+\omega_{\text{co}}^{2}}+\omega_{\text{co}}}{\sqrt{\left|\delta\right|^{2}+\omega_{\text{co}}^{2}}-\omega_{\text{co}}}\right)}{2\sqrt{\widetilde{\mu}}}\\
 & -\frac{\left|\delta\right|^{2}}{2\sqrt{\widetilde{\mu}}}\ln\left(\frac{\sqrt{\frac{\widetilde{\mu}+\omega_{\text{co}}}{\widetilde{\mu}}}+1}{\sqrt{\frac{\widetilde{\mu}+\omega_{\text{co}}}{\widetilde{\mu}}}-1}\right).
\end{align}
Up to second order in $\omega_{\text{co}}$, we find
\begin{equation}
\int_{0}^{\infty}d\omega\frac{1}{\sqrt{\omega}}\left[\omega-\widetilde{\mu}-\sqrt{\left(\omega-\widetilde{\mu}\right)^{2}+\left|\delta\right|^{2}}\right]\approx-\frac{\frac{1}{3}\left(4\widetilde{\mu}\right)^{2}+\left|\delta\right|^{2}\left(\ln\left[3\frac{\left(4\widetilde{\mu}\right)^{2}}{\left|\delta\right|^{2}}\right]+1\right)}{\sqrt{4\widetilde{\mu}}}.
\end{equation}
Now, we want to understand how this behaves as a function of $V_{2}$.
To this end, we reinsert $\widetilde{\mu}=\mu-eV_{2}$ and expand
up to second order in $V_{2}$. We get for the leading terms
\begin{equation}\label{eq_epsilon_charge_transistor}
\epsilon=-\frac{\frac{1}{3}\left(4\mu\right)^{2}+\left|\delta\right|^{2}\left(\ln\left[3\frac{\left(4\mu\right)^{2}}{\left|\delta\right|^{2}}\right]+1\right)}{\sqrt{4E_{L}\mu}}+4\sqrt{\frac{\mu}{E_{L}}}eV_{2}-\frac{1}{\sqrt{E_{L}\mu}}\left(eV_{2}\right)^{2},
\end{equation}
where we introduced the energy scale
\begin{equation}
E_{L}=\frac{\left(\hbar\frac{2\pi}{L}\right)^{2}}{2m_{e}}.
\end{equation}
There is a negative capacitance effect (the prefactor in front
of the $V_{2}^{2}$ term is negative), which however goes to zero in the thermodynamic limit. The linear term is an irrelevant rescaling of the total charge on node $2$. Consequently, we can discard the $V_2$-dependent terms in Eq.~\eqref{eq_epsilon_charge_transistor}. To reexpress the above expression in terms of more convenient quantities, we notice that the total number
of electrons can be related to $\mu$ as
\begin{equation}
\frac{\hbar^{2}\left(\frac{\pi}{L}\frac{N_{\text{tot}}}{2}\right)^{2}}{2m_{e}}=\mu.
\end{equation}
Consequently, we get
\begin{equation}
N_{\text{tot}}=4\sqrt{\frac{\mu}{E_{L}}}.
\end{equation}
We continue by defining $k_{F}$ as
\begin{equation}
k_{F}=\frac{\pi}{L}\frac{N_{\text{tot}}}{2}
\end{equation}
and consequently, $v_{F}$ is related to it by
\begin{equation}
v_{F}=\frac{\hbar k_{F}}{m_{e}}=\frac{\hbar}{m_{e}}\frac{\pi}{L}\frac{N_{\text{tot}}}{2}=\sqrt{2\mu/m_{e}}.
\end{equation}
Consequently, we find
\begin{equation}
\sqrt{E_{L}\mu}=\hbar\frac{\pi}{L}v_{F}=E_{\text{Th}},
\end{equation}
where $E_\text{Th}$ is the Thouless energy. We now see that for large $\mu$, if we define $N_2$ such that $eV_2=E_\text{Th}N_2$, a change of the quantum dot charge by a Cooper pair (corresponding to a change of its energy by $E_\text{Th}$) is equivalent to change of $N_2$ by $\pm 1$. Thus, $N_2$ is defined consistently for both transistor models.

Let us also here evaluate the leading deviation from a flat $\epsilon$ for a strongly asymmetric system, i.e., either $\Gamma_0\gg\Gamma_1$ or $\Gamma_0\ll\Gamma_1$. In the former case, we get (assuming $\mu\gg\Gamma$)
\begin{equation}\label{eq_approx_0_charge}
    \epsilon\approx \text{const.}-\frac{\Gamma_0}{2E_\text{Th}}\ln\left(\frac{\mu}{\Gamma_0}\right)\Gamma_1\cos(\phi_1)\ ,
\end{equation}
and for the latter case, we have to swap $\Gamma_0\leftrightarrow \Gamma_1$. We again observe that the deviation from a flat $\epsilon$ scales linearly with the weaker junction. Note that while $\mu$ tends to infinity in the thermodynamic limit, the natural logarithm ensures that the above expression increases very slowly and remains well-defined.

We proceed to computing the Berry curvature term. For the many-body
ground state, it is given as
\begin{equation}
\mathcal{B}=-2\sum_{n}\text{Im}\left[\partial_{\phi_{1}}u_{n-}^{*}\partial_{N_{2}}u_{n-}+\partial_{\phi_{1}}v_{n-}^{*}\partial_{N_{2}}v_{n-}\right].
\end{equation}
Let us reformulate this expression. First of all, we again take the
continuum limit
\begin{equation}
\mathcal{B}=-2\frac{1}{\sqrt{E_{L}}}\int_{0}^{\infty}d\omega\frac{1}{\sqrt{\omega}}\text{Im}\left[\partial_{\phi_{1}}u_{-}^{*}\left(\omega\right)\partial_{N_{2}}u_{-}\left(\omega\right)+\partial_{\phi_{1}}v_{-}^{*}\left(\omega\right)\partial_{N_{2}}v_{-}\left(\omega\right)\right],
\end{equation}
with
\begin{equation}
\left(\begin{array}{c}
u_{\pm}\left(\omega\right)\\
v_{\pm}\left(\omega\right)
\end{array}\right)=\frac{1}{\sqrt{2}}\left(\begin{array}{c}
\sqrt{1\pm\frac{\omega-\widetilde{\mu}}{\sqrt{\left[\omega-\widetilde{\mu}\right]^{2}+\left|\delta\right|^{2}}}}\\
\pm\frac{\delta}{\left|\delta\right|}\sqrt{1\mp\frac{\omega-\widetilde{\mu}}{\sqrt{\left[\omega-\widetilde{\mu}\right]^{2}+\left|\delta\right|^{2}}}}
\end{array}\right)\ ,
\end{equation}
using the same notation as above. Now, we use the fact that $u$ is real, and insert the explicit form
for $v$
\begin{equation}
\mathcal{B}=-\frac{1}{\sqrt{E_{L}}}\int_{0}^{\infty}d\omega\frac{1}{\sqrt{\omega}}\sqrt{1+\frac{\omega-\widetilde{\mu}}{\sqrt{\left[\omega-\widetilde{\mu}\right]^{2}+\left|\delta\right|^{2}}}}\partial_{N_{2}}\sqrt{1+\frac{\omega-\widetilde{\mu}}{\sqrt{\left[\omega-\widetilde{\mu}\right]^{2}+\left|\delta\right|^{2}}}}\text{Im}\left[\frac{\delta}{\left|\delta\right|}\partial_{\phi_{1}}\frac{\delta^{*}}{\left|\delta\right|}\right].
\end{equation}
This is further simplified as follows,
\begin{align*}
 & \sqrt{1+\frac{\omega-\widetilde{\mu}}{\sqrt{\left[\omega-\widetilde{\mu}\right]^{2}+\left|\delta\right|^{2}}}}\partial_{N_{2}}\sqrt{1+\frac{\omega-\widetilde{\mu}}{\sqrt{\left[\omega-\widetilde{\mu}\right]^{2}+\left|\delta\right|^{2}}}}\\
= & -E_\text{Th}\sqrt{1+\frac{\omega-\widetilde{\mu}}{\sqrt{\left[\omega-\widetilde{\mu}\right]^{2}+\left|\delta\right|^{2}}}}\partial_{\widetilde{\mu}}\sqrt{1+\frac{\omega-\widetilde{\mu}}{\sqrt{\left[\omega-\widetilde{\mu}\right]^{2}+\left|\delta\right|^{2}}}}\\
= & E_\text{Th}\frac{\left|\delta\right|^{2}}{2\left(\left[\omega-\widetilde{\mu}\right]^{2}+\left|\delta\right|^{2}\right)^{\frac{3}{2}}}.
\end{align*}
Consequently, we can already perform the integral, and get (again
under the assumption $\mu\gg\delta$)
\begin{equation}
\mathcal{B}\approx-\frac{E_\text{Th}}{\sqrt{\widetilde{\mu}E_{L}}}\text{Im}\left[\frac{\delta}{\left|\delta\right|}\partial_{\phi_{1}}\frac{\delta^{*}}{\left|\delta\right|}\right]\overset{\text{large \ensuremath{\mu}}}{\approx}-\text{Im}\left[\frac{\delta}{\left|\delta\right|}\partial_{\phi_{1}}\frac{\delta^{*}}{\left|\delta\right|}\right].
\end{equation}
Finally, we compute
\begin{align*}
\text{Im}\left[\frac{\delta}{\left|\delta\right|}\partial_{\phi_{1}}\frac{\delta^{*}}{\left|\delta\right|}\right] & =\text{Im}\left[\frac{\frac{\Gamma_0+\Gamma_1e^{i\phi_{1}}}{2}}{\left|\frac{\Gamma_0+\Gamma_1e^{i\phi_{1}}}{2}\right|}\partial_{\phi_{1}}\frac{\frac{\Gamma_0+\Gamma_1e^{-i\phi_{1}}}{2}}{\left|\frac{\Gamma_0+\Gamma_1e^{i\phi_{1}}}{2}\right|}\right]\\
 & =\text{Im}\left[\frac{\Gamma_0+\Gamma_1e^{i\phi_{1}}}{\sqrt{\Gamma_0^{2}+\Gamma_1^{2}+2\Gamma_0\Gamma_1\cos\left(\phi_{1}\right)}}\partial_{\phi_{1}}\frac{\Gamma_0+\Gamma_1e^{-i\phi_{1}}}{\sqrt{\Gamma_0^{2}+\Gamma_1^{2}+2\Gamma_0\Gamma_1\cos\left(\phi_{1}\right)}}\right]\\
 & =\text{Im}\left[-i\frac{\Gamma_1^{2}+\Gamma_0\Gamma_1\cos\left(\phi_{1}\right)}{\Gamma_0^{2}+\Gamma_1^{2}+2\Gamma_0\Gamma_1\cos\left(\phi_{1}\right)}\right]\\
 & =-\frac{\Gamma_1^{2}+\Gamma_0\Gamma_1\cos\left(\phi_{1}\right)}{\Gamma_0^{2}+\Gamma_1^{2}+2\Gamma_0\Gamma_1\cos\left(\phi_{1}\right)}.
\end{align*}
Consequently, we find
\begin{equation}
\mathcal{B}\approx\frac{\Gamma_1^{2}+\Gamma_0\Gamma_1\cos\left(\phi_{1}\right)}{\Gamma_0^{2}+\Gamma_1^{2}+2\Gamma_0\Gamma_1\cos\left(\phi_{1}\right)}\ ,
\end{equation}
which is in fact the exact same result as Eq.~\eqref{eq_Berry_charge_explicit}, when replacing $e_{J0,J1}\rightarrow \Gamma_{0,1}$. Consequently, the approximation for $\Gamma_0\gg\Gamma_1$ and $\Gamma_0\ll\Gamma_1$ is the same as in Eqs.~\eqref{eq_B_approx_0} and~\eqref{eq_B_approx_1}.


\section{Schrieffer-Wolff approximation for weakly nonflat energies and Berry curvatures}

In the main text, we discuss the impact of a deviation from flat bands, $\epsilon=\epsilon_0+\delta\epsilon(\phi_1)$ and $\mathcal{B}=\mathcal{C}+\delta\mathcal{B}(\phi_1)$. In this section, we show how to derive Eq.~\eqref{eq:leakage} in the main text by means of a standard Schrieffer-Wolff transformation, when assuming large $E_{L_2}$. To this end, we start with Eq.~\eqref{eq:H_cylinder}, and explicitly plug in the weak modulations $\delta\epsilon$ and $\delta\mathcal{B}$ in what we refer to as the Landau gauge in the main text, $\mathcal{A}_1=[\mathcal{C}+\delta\mathcal{B}(\phi_1)]N_2$ and $\mathcal{A}_2=0$,
\begin{equation}
    H=E_{C_1}\left[\widehat{N}_1+N_{g1}+\mathcal{C}\widehat{N}_2+\delta\mathcal{B}\left(\widehat{\phi}_1\right)\widehat{N}_2\right]^2+ E_{L_2}\widehat{\phi}_2^2 +E_{C_2}\widehat{N}_2^2+\delta\epsilon\left(\widehat{\phi}_1\right)\ ,
\end{equation}
where we discarded the irrelevant constant $\epsilon_0$. We can regroup the above Hamiltonian into three parts, $H=H_1+V+H_2$, where
\begin{equation}
    H_1=E_{C_1}\left(\widehat{N}_1+N_{g1}\right)^2+\delta\epsilon\left(\widehat{\phi}_1\right)\ ,
\end{equation}
and
\begin{equation}
    H_2=\left(E_{C_1}\mathcal{C}^2+E_{C_2}\right)\widehat{N}_2^2+ E_{L_2}\widehat{\phi}_2^2\ ,
\end{equation}
represent the physics of the separate degrees of freedom of node $1$ and $2$, respectively, whereas
\begin{equation}
    V=2E_{C_1}\left(\widehat{N}_1+N_{g1}\right)\mathcal{C}\widehat{N}_2+E_{C_1}\left\{\widehat{N}_1+N_{g1},\delta\mathcal{B}\left(\widehat{\phi}_1\right)\right\}\widehat{N}_2+2E_{C_1}\mathcal{C}\delta\mathcal{B}\left(\widehat{\phi}_1\right)\widehat{N}_2^2+\mathcal{O}\left[\delta\mathcal{B}^2\right]\ ,
\end{equation}
describes their coupling up to first order in $\delta\mathcal{B}$ (the anticommutator $\{\cdot,\cdot\}$ is necessary in the second term, because $N_1$ obviously does not commute with $\delta\mathcal{B}$).

The eigenmode of $H_2=\omega_\text{LC}\widehat{b}^\dagger\widehat{b}+\text{const.}$ is the LC resonator mode, whose frequency $\omega_\text{LC}$ and ladder operator $\widehat{b}$ have already been given in the main text. For the Schrieffer-Wolff transformation, we project onto the LC resonator ground state, $\vert 0_\text{LC}\rangle$, satisfying $\widehat{b}\vert 0_\text{LC}\rangle=0$. Defining the projector as $P=\vert 0_\text{LC}\rangle\langle 0_\text{LC}\vert$, the resulting low-energy Hamiltonian is
\begin{equation}
    H\approx H_1+PVP-PV\frac{1}{H_2}(1-P)VP\ ,
\end{equation}
(note that $PH_2P$ is by definition an irrelevant constant) which yields up to first order in $\delta\mathcal{B}$
\begin{equation}
    H\approx E_{C,\text{eff}}(\mathcal{C})\left[\left(\widehat{N}_1+N_{g1}\right)^2+\delta\epsilon-\frac{\mathcal{C}}{2}\frac{\omega_\text{LC}}{E_{C_2}}\delta\mathcal{B}-\frac{\mathcal{C}}{2}\frac{E_{C_1}}{E_{C_2}}\{\widehat{N}_1+N_{g1},\widehat{N}_1+N_{g1},\delta\mathcal{B}\}\}\right]\ .
\end{equation}
The second and third term are pure Josephson leakage currents, whereas the fourth term is a more exotic mixed term, containing both charge and phase operators. However, due to $E_{L_2}$ large (and thus $\omega_\text{LC}$ large), it is always inferior to the third term, and thus nominally a higher order effect. Neglecting it, we arrive at Eq.~\eqref{eq:leakage} in the main text.

In the above supplementary sections, we have provided explicit expressions for $\delta\epsilon$ and $\delta\mathcal{B}$ for the two concrete transistor models, see Eqs.~(\ref{eq_eps_approx_0}-\ref{eq_B_approx_1}) and~\eqref{eq_approx_0_charge}. Here, both types of corrections yield ordinary sequential Cooper pair tunneling processes similar to a regular SIS Josephson junction, $\sim\cos(\phi_1)$.

\end{document}